\documentclass[journal]{IEEEtran}
 \usepackage{amsmath,amssymb}
 \usepackage{subfigure}
 \usepackage{graphicx,graphics,color,psfrag}
 \usepackage{cite,balance}
 \usepackage{algorithm}
 \usepackage{accents}
 \usepackage{amsthm}
 \usepackage{bm}
 \usepackage{url}
 \usepackage{algorithmic}
 \usepackage[english]{babel}
 \usepackage{multirow}
 \usepackage{enumerate}
 \usepackage{cases}
 \usepackage{stfloats}
 \usepackage{dsfont}
 \usepackage{color,soul}
 \usepackage{amsfonts}
 \usepackage{cite,graphicx,amsmath,amssymb}
 \usepackage{fancyhdr}
 \usepackage{hhline}
 \usepackage{graphicx,graphics}
 \usepackage{array,color}
 \usepackage{amsmath}
 \usepackage{amsthm}

\newtheorem{remark}{Remark}
\newtheorem{theorem}{Theorem}
\newtheorem{lemma}{Lemma}

\include{header}

\IEEEoverridecommandlockouts

\begin{document}

\title{Optimized Power Control for Over-the-Air Computation in Fading Channels}
\author{Xiaowen Cao, Guangxu Zhu, Jie Xu, and Kaibin Huang \\
\thanks{Manuscript received June 18, 2019; revised November 27, 2019 and April 20, 2020; accepted July 20, 2020. This work was supported in part by the Key Area R\&D Program of Guangdong Province with grant No. 2018B030338001, the National Key R\&D Program of China with grant No. 2018YFB1800800, the Natural Science Foundation of China with grant No. 61871137,  the Guangdong Province Basic Research Program (Natural Science) with grant No. 2018KZDXM028, the Hong Kong Research Grants Council with grants No. 17208319 and No. 17209917, the Innovation and Technology Fund with grant No. GHP/016/18GD, the Guangdong Basic and Applied Basic Research Foundation with grant No. 2019B1515130003, and the Shenzhen Peacock Plan with grant No. KQTD2015033114415450. Part of this paper has been presented at the IEEE Global Communications Conference (GLOBECOM), Waikoloa, HI, USA, Dec.9--13, 2019.}
\thanks{X. Cao is with the School of Information Engineering, Guangdong University of Technology, Guangzhou, China, and the Future Network of Intelligence Institute (FNii), The Chinese University of Hong Kong (Shenzhen), Shenzhen, China (e-mail: caoxwen@outlook.com).}
\thanks{J. Xu is with the Future Network of Intelligence Institute (FNii) and the School of Science and Engineering (SSE), The Chinese University of Hong Kong (Shenzhen), Shenzhen, China (xujie@cuhk.edu.cn).  J.~Xu is the corresponding author.}
\thanks{G. Zhu is with Shenzhen Research Institute of Big Data, Shenzhen, China  (e-mail: gxzhu@sribd.cn). }
\thanks{K. Huang is with the Dept. of Electrical and Electronic Engineering, The University of Hong Kong, Pok Fu Lam, Hong Kong (e-mail: huangkb@eee.hku.hk). }
}

\markboth{}{}
\maketitle

\setlength\abovedisplayskip{2pt}
\setlength\belowdisplayskip{2pt}

 \begin{abstract}
  \emph{Over-the-air computation} (AirComp) of a function (e.g., averaging) has recently emerged as an efficient multiple-access scheme for fast aggregation of distributed data at mobile devices (e.g., sensors) at a \emph{fusion center} (FC) over wireless channels. To realize reliable AirComp in practice, it is crucial to adaptively control the devices'  transmit power for coping with channel distortion to achieve the desired magnitude alignment of simultaneous signals.
  In this paper, we solve the power control problem. Our objective is to minimize the computation error by jointly optimizing the transmit power at devices and a signal scaling factor (called denoising factor) at the FC, subject to individual average power constraints at devices.
  The problem is generally non-convex due to the coupling of the transmit powers at devices and denoising factor at the FC.
  To tackle the challenge, we first consider the special case with static channels, for which we derive the optimal solution in closed form.
  The derived power control exhibits a \emph{threshold-based} structure: if the product of the channel quality and power budget for each device, called quality indicator, exceeds an optimized threshold, this device applies channel-inversion power control; otherwise, it performs full power transmission.
  We proceed to consider the general case with time-varying channels. To solve the more challenging non-convex power control problem, we use the Lagrange-duality method via exploiting its ``time-sharing" property.
  The derived power control exhibits a \emph{regularized channel inversion} structure, where the regularization balances the tradeoff between the signal-magnitude alignment and noise suppression. Moreover, for the special case with only one device being power limited, we show that the power control for the power-limited device has an interesting \emph{channel-inversion water-filling} structure, while those for other devices (with sufficiently large power budgets) reduce to \emph{channel-inversion} power control. Numerical results show that the derived power control significantly reduces the computation error as compared with the conventional designs.
\end{abstract}

\vspace{-0.5cm}

\section{Introduction}\label{sec:intro}
Future {\em Internet-of-Things} (IoT) networks are envisioned to comprise an enormous number of mobile devices (e.g., sensors) for enabling big-data applications such as smart cities and realizing edge {\it artificial intelligence} (AI) \cite{5Gsurvey,Chiang16}. Therefore, how to efficiently aggregate massive data distributed at dense devices is becoming increasingly important.
A technology called {\em over-the-air computation} (AirComp) has recently emerged as a promising multi-access scheme for such fast {\em wireless data aggregation} (WDA), especially in ultra-low-latency and high-mobility scenarios (e.g., for patrolling drones or emergency medical devices) \cite{nomo_function_Abari}.
In AirComp, a {\it fusion center} (FC) can exploit the signal-superposition property of a {\it multiple-access channel} (MAC), for efficiently averaging the simultaneously transmitted data by many devices in an analog manner.
With proper pre-processing at devices and post-processing at the FC, AirComp can perform not just averaging but compute a class of so-called {\it nomographic functions} of distributed data such as geometric mean and polynomial. Recently, the applications of  AirComp have expanded from sensor networks towards networks supporting edge machine learning \cite{GX18_learning,Kang19}.

The idea of AirComp was first proposed in the pioneering work \cite{nomo_function_Nazer}, where structured codes were designed for reliable functional computation over a MAC by exploiting the interference caused by simultaneous transmissions. Building upon this work, it was proved that the simple uncoded analog transmission is optimal in achieving the minimum distortion in a network with {\it independent and identically distributed} (i.i.d.) Gaussian data sources \cite{Gastpar08}.
However, coding is required for distortion minimization in other scenarios with, e.g., bivariate Gaussian distributed \cite{Wagner08} and correlated Gaussian distributed data sources \cite{Soundararajan12}.
In another line of research, AirComp with uncoded analog transmission was investigated from the signal processing perspective. For instance, in wireless sensor networks, linear decentralized estimation with static channels was studied in \cite{Xiao08}, while, the optimal distortion outage performance for AirComp was investigated in \cite{Wang11}, where an outage happens when the estimation error or distortion exceeds a predetermined threshold.
A so-called ``AirShare" design in \cite{Abari15} was presented for resolving the synchronization problem in distributed transmission, in which the FC broadcasts a shared clock to all devices.
More recently, the {\it multiple-input-multiple-output} (MIMO) technique was explored in \cite{GX18,DZ18} to enable vector-valued AirComp targeting multi-modal sensing, in which zero-forcing precoding at sensors and aggregation beamforming at FC were jointly designed for minimizing the computation error measured by the {\it mean squared error} (MSE).
Along this vein, the authors in \cite{XY18} integrated MIMO AirComp with {\it wireless power transfer} (WPT) to enable self-sustainable AirComp for low-power devices.
Besides, a blind MIMO AirComp without requiring {\it channel state information} (CSI) was proposed in \cite{ZD18_blind} for low-complexity and low-latency IoT networks, while an intelligent reflecting surface assisted AirComp system was investigated in \cite{TJ18_IRS}.
In addition, AirComp also finds applications beyond distributed sensing in wireless sensor networks. For example, a compute-and-forward scheme for wireless relaying systems was presented in \cite{Nazer11}, in which based on a similar idea like AirComp, the relay computes linear functions of messages sent from sources and forwards them to destinations for robust decoding.
Moreover, the authors in \cite{GX18_learning,Kang19} applied AirComp in the federated learning system to enable communication-efficient collaborative AI-model training by leveraging distributed data at edge devices.

To enable reliable AirComp over fading channels, it is crucial to adapt the devices' transmit power to channel states for coping with channel distortion caused by noise and fading, thereby achieving the desired signal-magnitude alignment at the FC.
Despite extensive research, most prior work on AirComp assumes the simple channel-inversion power control (or equivalently zero-forcing precoding in the MIMO case) at devices, such that their transmitted signals are perfectly aligned in magnitude at the FC receiver to yield the desired function \cite{GX18,DZ18,XY18,GX18_learning,Kang19,TJ18_IRS}.
However, the scheme is sub-optimal and can severely degrade the AirComp performance in the presence of deep fading.
Specifically, it is well known that channel inversion may significantly enhance noise, especially when devices are subject to stringent power constraints, a typical case for sensors.
It has been well established that optimal power control over fading channels is crucial to approach the performance limit of wireless communication systems \cite{Goldsmith} (e.g., for achieving the capacity in point-to-point fading channels \cite{Goldsmith1997,Biglieri1998} and cognitive radio systems
\cite{Ghasemi2007,Kang2009,RZ09}).
For the same reason, it is equally important to investigate optimal power control for AirComp over fading channels.
This topic, however, is a largely uncharted area. Although there exist relevant studies on the optimal power control to minimize the MSE for decentralized estimation \cite{Xiao08,Wang11}, they focus on the static channel scenario with a total power constraint for all devices. The scenario does not fit the AirComp system of our interest. This motivates the current work.

In this paper, we consider an IoT network with AirComp over a fading MAC, where mobile devices simultaneously transmit the sensing data to the FC for averaging.
We investigate the power control problem for reliable AirComp, where a general data source model with possibly correlated observations between devices is adopted.
Our objective is to minimize the computation error (measured by MSE)  by jointly optimizing the transmit power at each device and the signal scaling factor for noise suppression, called denoising factor, applied at the FC, subject to devices' individual average power constraints.
Nevertheless, the MSE minimization requires optimally balancing the tradeoff between reducing the error in signal alignment (required for averaging) and suppressing the noise.
This also leads to a challenging optimization problem due to its non-convexity arising from the coupling of the control variables, namely the transmit powers and denoising factor.
Through the joint design, the derived power control contributes a new dimension for reducing the computation error in AirComp.
The main contributions of this work are summarized as follows.
\begin{itemize}
	\item { \bf Power control for static channels:} In this case, the channels are fixed in time but vary over devices.
	The original problem is reduced to a non-convex optimization problem. The derived power-control policy in closed form is found to have a \emph{threshold-based} structure. The threshold is applied on a derived {\it quality indicator} at each device which accounts for both the power budget of the device and its channel power gain. It is shown that if the quality indicator exceeds the optimized threshold, the device applies channel-inversion power control; otherwise, it performs full power transmission. Furthermore, asymptotic analysis is provided to show the behaviour of the optimal policy for high and low \emph{signal-to-noise ratio} (SNR) regimes, corresponding to channel inversion and full power transmission, respectively.
	\item {  \bf Power control for time-varying channels:}  In this case, the channels vary both in time and over devices. Due to the coupling of the power control at devices and time-varying denoising factor, the formulated problem is a more challenging non-convex problem.
	But the problem is shown to satisfy the ``time-sharing" condition, and thus the strong duality holds between the original problem and its dual problem. This motivates us to leverage the Lagrange-duality method to solve the problem.
	As a result, the optimal power-control policy over devices and fading states is derived to have the structure of \emph{regularized channel inversion}, where the regularization has the function of balancing the signal-magnitude alignment and noise suppression.
	Moreover, for the special case with only one device being power limited, we show that the optimal power control policy for this device is to combine \emph{channel-inversion} and \emph{water-filling}, while that for other devices (with sufficiently large power budgets) to perform channel inversion.
   \item  {\bf Simulation:}
   Simulation results are presented to validate the derived analytical results. It is shown that the derived power control policy substantially improves the AirComp performance, compared with conventianal designs such as uniform or channel-inversion power control.
    Furthermore, it is observed that the performance gain over the uniform power control is marginal in the low SNR regime but becomes substantial as the SNR increases. This finding is in sharp contrast with the conventional water-filling power control for rate maximization targeting communication over fading channels (see, e.g., \cite{Goldsmith1997}), where the adaptive power control is crucial in the low to moderate SNR regimes.
\end{itemize}

The remainder of the paper is organized as follows. The fading AirComp system is modeled in Section \ref{sec:system1}, where the power control problem for MSE minimization is formulated. The optimal power-control policy in the special static-channel case is presented in Section \ref{sec_static_channel}.
The optimal power-control policy in the general case with time-varying channels is obtained in Section \ref{sec_fading}. Finally, simulation results are provided in Section \ref{sec_simu}, followed by the conclusion in Section \ref{sec_con}.

\section{System Model and Problem Formulation}\label{sec:system1}
\begin{figure}
\centering
 \setlength{\abovecaptionskip}{-4mm}
\setlength{\belowcaptionskip}{-4mm}
    \includegraphics[width=8cm]{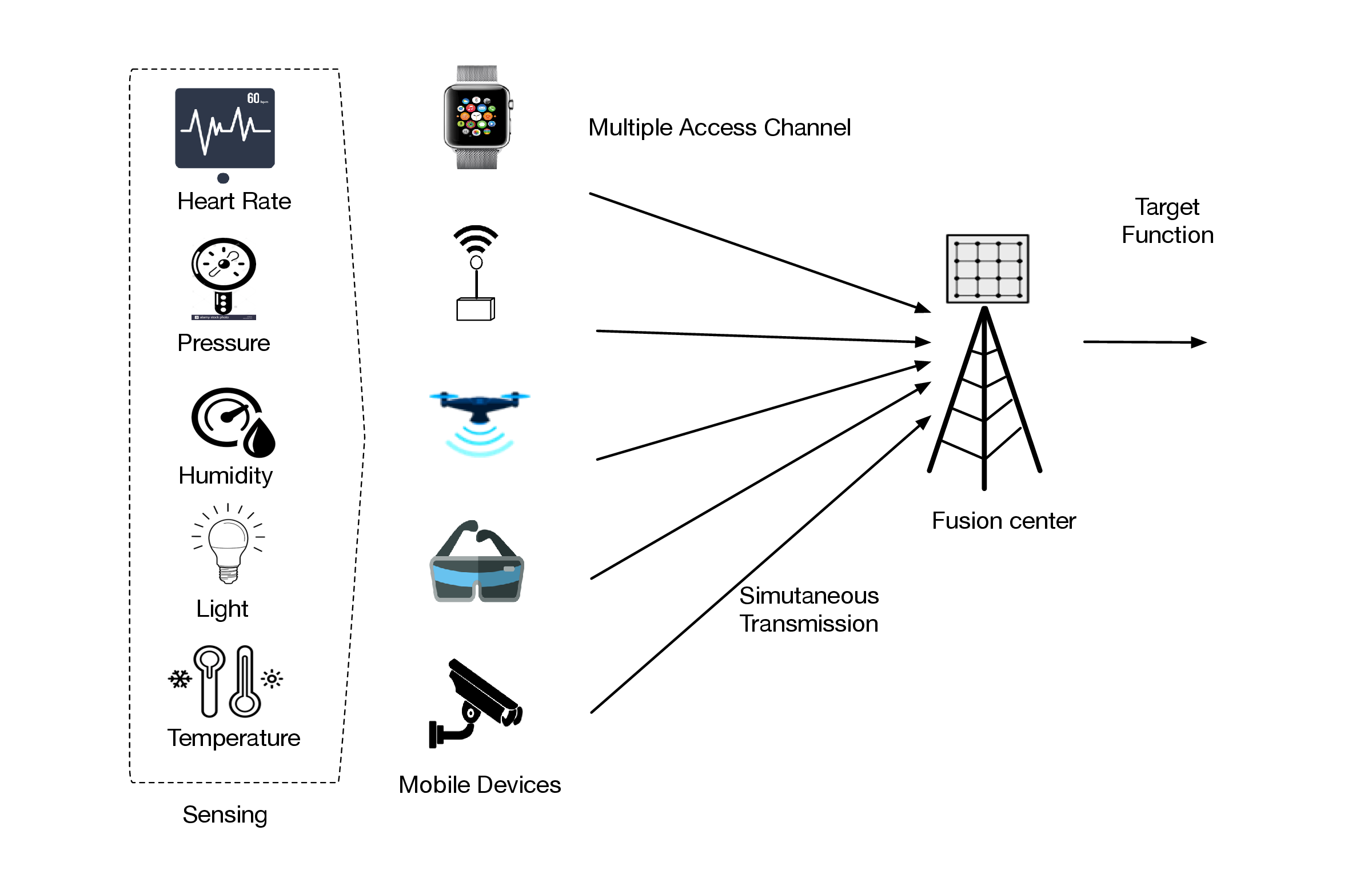}
\caption{The AirComp system for fast wireless data aggregation.} \label{fig:model}
\vspace{-0.1cm}
\end{figure}

We consider AirComp over a MAC as shown in Fig. \ref{fig:model}, in which one FC aggregates the information from a set $\mathcal {K}\triangleq \{1,\ldots,K\} $ of $K\ge 1$ mobile devices each with one antenna.
We consider the case with block fading channels, where the wireless channels remain unchanged within each time slot but may change from one slot to another. Let $\mathcal T \triangleq \{1, \ldots , T\}$ denote the set of time slots of interest, where $T$ is the number of slots that is assumed to be sufficiently large. It is assumed that the channel coefficients over different time slots are generated from a stationary and ergodic stochastic process.
Suppose that the devices record a set of time-varying parameters of an environment, and the FC needs to recover the \emph{average} of the measured data from devices\footnote{Other functions such as the geometric mean can also be computed via proper pre-processing and post-processing \cite{nomo_function_Abari}.}.

Thus the function of interest at the FC in time slot $t$ is \begin{align}
\tilde f(t)=\frac{1}{K}\sum \limits_{k\in{\cal K}}Z_{k}(t).
\end{align}
Instead of directly transmitting $Z_{k}(t)$, we find it convenient to transmit its normalized version, denoted by $s_{k}(t)=g(Z_{k}(t))$, to facilitate power control. The function $g(\cdot)$ denotes the normalization operation, which is linear and uniform to all devices, to ensure that $\{s_{k}(t)\}$ are with zero mean and unit variance.
Upon receiving the average of transmitted data $\{s_k(t)\}$, i.e.,
\begin{align}
f(t)=\frac{1}{K} \sum \limits_{k\in{\cal K}}s_{k}(t),\label{functions}
\end{align}
the desired $\tilde f(t)$ can be simply recovered from $f(t)$ by applying the following de-normalization operation:
\begin{align}
\tilde f(t) =g^{-1}(f(t)),
\end{align}
where $g^{-1}(\cdot)$ denotes the inverse function of $g(\cdot)$.
Due to the one-to-one mapping between $f(t)$ and $\tilde f(t)$, we, hereafter, refer to $f(t)$ as the target-function value in slot $t$ for ease of exposition.

In an arbitrary time slot $t$, let $h_k(t)$ and $b_k(t)$ denote the channel coefficient from device $k$ to the FC and the transmit coefficient at device $k$, respectively.
The received signal at the FC is
\begin{align}
y(t)=\sum \limits_{k\in{\cal K}}h_{k}(t)b_k(t)s_{k}(t)+ w(t),
\end{align}
where $w(t)$ denotes the {\em additive white Gaussian noise} (AWGN) at the FC receiver with zero mean and variance of $\sigma^2$.
Assume $b_k(t)\!=\!\frac{ \sqrt{p_{k}(t)} h_{k}^{\dagger}(t)}{ |h_{k}(t)|}$, where $p_k(t)\!\ge \!0$ denotes the transmit power at device $k\in\! \cal K$ in each time slot $t$ and ${\dagger}$ denotes the conjugate operation.
 Hence, we have\footnote{For simplicity, we assume the availability of perfect CSI at each device transmitter and the FC receiver. Therefore, the time-varying phase shifts introduced by channel fading can be perfectly compensated, which allows us to focus on the transmit power control only. In practice, the CSI can be obtained at each device via channel estimation based on the downlink pilots transmitted over the same frequency band from the FC, by using the uplink-downlink channel reciprocity. }
\begin{align}
y(t)=\sum \limits_{k\in{\cal K}}\sqrt{p_{k}(t)}|h_{k}(t)| s_{k}(t)+ w(t). \label{time_Yv}
\end{align}
Upon receiving the signal $y(t)$ in \eqref{time_Yv}, the FC applies the denoising factor denoted by $\eta(t)$, to recover the average message of interest as
\begin{align}\label{ave_functions}
\hat f(t) = \frac{y(t)}{K\sqrt{\eta(t)}},
\end{align}
where the post-processing $1/K$ is used in \eqref{ave_functions} for averaging. We are interested in minimizing the distortion of the recovered average of the transmitted data, \emph{with respect to} (w.r.t.) the ground truth average $f(t)$.	
In any given time slot $t$, the distortion is measured by the corresponding instantaneous MSE defined as
\begin{align}\label{MSE_ave}
 \widehat {\rm MSE}\!(t)\!&\!=\!\mathbb{E}\!\left[\!\left(\!\hat{f}\!(t)\!-\!f(t)\!\right)^2\!\right]\!=\!\frac{1}{K^2}\!\mathbb{E}\!\left[\!\left(\!\frac{y(t)}{\sqrt{\eta(t)}}\!-\! \!\sum \limits_{k\in{\cal K}}\!s_{k}\!(t)\!\right)^2\! \right]\!\notag\\
 &\!\!\!\!\!\!\!\!\!\!=\! \frac{1}{K^2}\!\mathbb{E}\!\!\left[\!\! \left(\!\sum \limits_{k\in{\cal K}}s_k\!(t)\!\! \left(\!\frac{\!\sqrt{p_{k}(t)}|h_{k}(t)|}{\sqrt{\eta(t)}}\!-\!1\!\right)\!+\!\frac{w(t)}{\sqrt{\eta(t)}}\!\!\right)^2 \!\right],\!
\end{align}
where the expectation is taken over the distribution of the transmitted signals $\{s_k(t)\}$.
In general, the sensing observations $\{s_k(t)\}$ can either be statistically independent or correlated among different devices.
In the case when $\{s_k(t)\}$ are statistically independent among different users (i.e.,$ \mathbb{E}[s_i(t)s_j(t)] = 0, \forall i\neq j$), the instantaneous MSE is
\begin{align}\label{MSE_t2}
\!\!\!\! \widehat{\rm MSE}\!(t)\!\!=\!\!{\rm MSE}\!(t)\!\triangleq \!\!\frac{1}{K^2}\!\!\left(\!\sum \limits_{k\in{\cal K}}\!\!\!\left(\!\frac{\sqrt{p_{k}(t)}|h_{k}(t)|}{\sqrt{\eta(t)}}\!-\!1\!\right)^2\!\!\!+\!\!\frac{{\sigma}^2}{\eta(t)}\!\right).\!
\end{align}
Such independent sensing observations have been widely considered in the literature \cite{GX18,Kang19,Gastpar08,XY18,TJ18_IRS}.

In the more general case when  $\{s_k(t)\}$ are correlated among different devices (i.e., $\mathbb{E}[s_i(t)s_j(t)]=c_{ij}, \forall i\neq j$), the  instantaneous MSE is given as
\begin{align}\label{MSE_t1}
& \widehat {\rm MSE}(t)\notag\\
&\! \!=\!\! \!\frac{1}{K^2}\!\!\left(\!\!\sum \limits_{i\in{\cal K}}\!\sum \limits_{j\in{\cal K}}\!\!\left(\!\!\!\frac{\sqrt{p_{i}(t)}|h_{i}(t)|}{\sqrt{\eta(t)}}\!\!-\!\!1\!\!\right)\!\!\!\left(\!\!\!\frac{\sqrt{p_{j}(t)}|h_{j}(t)|}{\sqrt{\eta(t)}} \!-\!\!1\!\!\right)\!\!c_{ij}\!\!+\!\frac{\sigma^2}{\eta(t)}\!\!\right)\!.\!
\end{align}
However, the evaluation of $ \widehat {\rm MSE}(t)$ requires the perfect knowledge of the cross correlation $c_{ij}$ among different devices, which is difficult to obtain in practice. Moreover, the function is complicated due to the coupling between transmission power at difference devices.
As a result, the MSE derived from the generalized model is intractable for optimization. The above difficulty can be overcome by deriving a tractable upper bound given as
\begin{align}\label{MSE_t3}
& \widehat {\rm MSE}(t)\notag\\
 &\!= \!\frac{1}{K^2}\!\mathbb{E}\!\left[ \!\left(\!\sum \limits_{k\in{\cal K}}\!s_k(t) \!\left(\!\frac{\sqrt{p_{k}(t)}|h_{k}(t)|}{\sqrt{\eta(t)}}\!-1\right)\!+\frac{w(t)}{\sqrt{\eta(t)}}\!\right)^2\! \right]\notag\\
&\leq\!\! \frac{K\!+\!1}{K^2}\mathbb{E}\!\!\left[\!\sum \limits_{k\in{\cal K}}\!s_k^2(t)\! \left(\!\frac{\sqrt{p_{k}(t)}|h_{k}(t)|}{\sqrt{\eta(t)}}\!\!-\!1\!\!\right)^2\!\!\!\!+\!\!\left(\!\!\frac{w(t)}{\sqrt{\eta(t)}}\!\right)^2\!\right]\notag\\
&= (K+1){\rm MSE}(t) ,
\end{align}
where the inequality follows due to the Cauchy's inequality $(x_1y_1+x_2y_2+\cdots+x_n y_n)^2\leq(x_1^2+x_2^2+\cdots+x_n^2 )(y_1^2+y_2^2+\cdots+ y_n^2)$.
Therefore, for the objective of power control optimization, we can use the upper bound $(K+1){\rm MSE}(t)$ as an approximation of $ \widehat {\rm MSE}(t)$ when sensing observations are correlated, and use the exact $ \widehat {\rm MSE}(t)$ according to \eqref{MSE_t2} when sensing observations are statistically independent.
This leads to a unified optimization problem that can account for both cases (up to a difference in the scaling factor in the objective).

Based on \eqref{MSE_ave} and \eqref{MSE_t3}, for sufficiently large $T$ (or $T \to \infty$), the time-averaging MSE of interest is given by
\begin{align}
	\overline{\rm MSE} &=\! \lim_{T\to \infty} \frac{1}{T}\sum\limits_{t\in\cal T} {\rm MSE}(t)\notag\\
	&\!\!=\! \!\lim_{T\!\to\! \infty} \!\frac{1}{T}\!\sum\limits_{t\in\cal T}\!\frac{1}{K^2}\!\!\left(\!\!\sum \limits_{k\in{\cal K}}\!\!\left(\!\frac{\sqrt{p_{k}(t)}|h_{k}(t)|}{\sqrt{\eta(t)}}\!-\!1\!\right)^2\!\!\!\!\!+\!\!\frac{\sigma^2}{\eta(t)}\!\!\right),\!\!\label{MSE_s_y_t}
\end{align}
where the scaling factor ($K+1$) of ${\rm MSE} (t)$ for the correlated sensing case is omitted without any loss in optimality.
Note that the channel coefficients over slots are assumed to be generated from a stationary and ergodic stochastic process.
Let $\bm \nu\triangleq[h_1,\cdots,h_K]$ denote the channel vector comprising the wireless channel coefficients from the $K$ device transmitters to the FC receiver, or the fading state \cite{RZ09}.
Let $\psi(\bm\nu)$ denote the probability, or the fraction of time, when the channel is in the fading state $\bm\nu$. Accordingly, we define $\mathbb{E}_{\bm\nu}[\tilde{f}(\bm\nu)] \!\!=\!\! \int_{\bm\nu}\! \tilde{f}(\bm\nu) \psi(\bm\nu) d\bm\nu$ as the expectation of any function $\tilde{f}(\bm\nu)$ over fading distributions.
 By using the stationary and ergodic nature of the fading channels, the time-average MSE in \eqref{MSE_s_y_t} can be translated to the following ensemble-average MSE:
\begin{align}
	\overline{\rm MSE}&=  \mathbb{E}_{\bm \nu} \left[{\rm MSE}(t)\right] \notag\\
	&= \frac{1}{K^2} \mathbb{E}_{\bm \nu}\left[\sum \limits_{k\in{\cal K}}\left(\frac{\sqrt{p_{k}(\bm \nu)}|h_{k}|}{\sqrt{\eta(\bm \nu)}}-1\right)^2+\frac{\sigma^2}{\eta(\bm \nu)}\right],\label{MSE_s_y}
\end{align}
where $p_k(\bm \nu)$  and $\eta(\bm \nu)$ denote device $k$'s transmit power and the FC's denoising factor at any fading state $\bm \nu$, respectively.

Furthermore, in practice, each device $k \in \mathcal K$ is constrained by an average power budget $\bar P_k$. Therefore, we have the devices' individual average transmit power constraints as
\begin{align}
\mathbb{E}_{\bm\nu}[p_k(\bm\nu)]\leq \bar{P}_k,~\forall k\in\cal K.\label{fad_bar_P_ave}
\end{align}

Our objective is to minimize $\overline{\rm MSE}$ in \eqref{MSE_s_y}, by jointly optimizing the power control $\{ p_{k}(\bm\nu)\}$ at devices and the denoising factors $\{\eta(\bm\nu)\}$ at the FC, subject to the individual average transmit power constraints at devices in \eqref{fad_bar_P_ave}.
Therefore, the optimization problem of interest is formulated as (P1) in the following, where we omit the constant coefficient $\frac{1}{K^2}$ in \eqref{MSE_s_y} for notational convenience:
\begin{align}
\mathbf{(P1):}\min_{\substack{\{ p_k(\bm\nu)\ge 0\},\\
	  \{\eta(\bm\nu)\ge 0\}}} ~& \! \mathbb{E}_{\bm\nu}\left[\sum \limits_{k\in{\cal K}}\left(\frac{\sqrt{p_{k}(\bm\nu)}|h_{k}|}{\sqrt{\eta(\bm\nu)}}\!-\!1\right)^2\!+\!\frac{\sigma^2}{\eta(\bm\nu)}\right]\notag \\
{\rm s.t.}~~~~~~~~&\mathbb{E}_{\bm\nu}[p_k(\bm\nu)]\leq \bar{P}_k,~\forall k\in\cal K\notag.
\end{align}
It is observed that the objective function of problem (P1) (i.e., the ensemble-average MSE) consists of two components representing the signal misalignment error
(i.e., $\mathbb{E}_{\bm\nu}\!\!\left[\! \sum\limits_{k\in{\cal K}}\!\! \!\left(\!\!\frac{\sqrt{p_{k}(\bm\nu)}|h_{k}|}{\sqrt{\eta(\bm\nu)}}\!-\!1\!\!\right)^2\!\right]$) and the noise-induced error (i.e., $\mathbb{E}_{\bm\nu}[ \sigma^2/\eta(\bm\nu)]$), respectively.
In general, enlarging $\eta(\bm\nu)$ can reduce the noise-induced error component but lead to an increased signal misalignment error (due to limited transmit power at devices),
while reducing $\eta(\bm\nu)$ can suppress the signal misalignment error (as less power is required for signal alignment) but at the cost of enhancing the noise-induced error. Therefore, in problem (P1) of our interest, there exists a \emph{fundamental tradeoff} between minimizing the signal misalignment error and suppressing the noise-induced error. Moreover, due to the coupling of the transmit power $\{p_k(\bm\nu)\}$ and denoising factors $\{\eta(\bm\nu)\}$, problem (P1) is non-convex in general, and thus challenging to solve.

\vspace{-0.1cm}
\section{Power Control with Static Channels}\label{sec_static_channel}
In this section, we consider the special case of problem (P1) with static channels, namely $\bm \nu$ remains unchanged over time.
Accordingly, we suppose that different devices apply fixed transmit power and the FC applies a fixed denoising factor, i.e., $ p_{k}(\bm\nu)=p_{k}, \eta(\bm\nu)=\eta, \forall k\in{\cal K}$.
Furthermore, to facilitate the subsequent derivation, for each device $k\in \cal K$, we define the \emph{quality indicator} as the product of its power budget and channel power gain (i.e., $\bar{P}_k|h_k|^2$). Then we make the following assumption without loss of generality:
\begin{align}\label{static_order}
 \bar{P}_1|h_1|^2\leq \cdots\leq\bar{P}_k|h_k|^2\leq\cdots\leq \bar{P}_K|h_K|^2.
\end{align}
As a result, problem (P1) is reduced to the following power control problem: \begin{align*}
\mathbf{(P2):}\min_{\{ p_k\geq0\},\eta\geq0} ~& \sum \limits_{k\in\cal K}\left(\frac{\sqrt{p_{k}}|h_{k}|}{\sqrt{\eta}}-1\right)^2+\frac{\sigma^2}{\eta}\\
{\rm s.t.}~~~~& p_{k}\leq \bar{P}_k, ~\forall k\in\mathcal {K}.
\end{align*}
However, problem (P2) is still non-convex due to the coupling of $\{p_k\}$ and $\eta$.
To overcome the challenge, we first optimize over $\{p_k\}$ with any given $\eta \ge 0 $, and then search for the optimal $\eta$. The derived optimal solution exhibits an interesting threshold-based structure as elaborated in the following subsection.
\vspace{-0.5cm}
\subsection{Threshold-Based Power Control Policy}\label{static_optimal}
In this subsection, we first present the optimal solution to problem (P2) by optimizing over $\{p_k\}$ with any given $\eta\geq 0$, and then searching for the optimal $\eta$.
First, we consider the optimization over $\{p_k\}$ with a given $\eta \geq 0$. In this case, we decompose problem (P2) into the following $K$ subproblems each for optimizing $p_k$:
\begin{align}\label{static_p_k}
\min_{ 0\le p_k \le \bar P_k } ~&\left( \frac{\sqrt{p_{k}}|h_{k}|}{\sqrt{\eta}}-1\right)^2,
\end{align}
where the constant term $\sigma^2/\eta$ w.r.t. $p_k$ is ignored.
It is evident that the optimal power-control policy to problem \eqref{static_p_k} for device $k$ is given by
\begin{align}
p_k^{*}=\min~\left(\bar{P}_k,\frac{\eta}{|h_k|^2}\right).\label{K_device_N1_Pk_opyt}
\end{align}

Next, we proceed to optimize over $\eta$. By substituting $p_k^{*}$ in \eqref{K_device_N1_Pk_opyt} back to problem (P2), we have the optimization problem over $\eta$ as
 \begin{align}
\min_{\eta\geq 0} ~ F(\eta) \triangleq \sum \limits_{k\in{\cal K}}\left(\min\left(\frac{\sqrt{\bar{P}_k}|h_{k}|}{\sqrt{\eta}}-1,0\right)\right)^2+\frac{\sigma^2}{\eta},\label{static_K_eta}
\end{align}
with $F(\eta)$ denoting the objective function w.r.t. $\eta$.
To  solve problem \eqref{static_K_eta}, we need to remove the ``min'' operation to simplify the derivation. To this end, we find it convenient to adopt a \emph{divide-and-conquer} approach that divides the feasible set of problem \eqref{static_K_eta}, namely $\{\eta\geq 0\}$, into $K+1$ intervals. Note that each of them is defined as
\begin{align}
	{\cal F}_k\!=\!\{\eta \mid \bar{P}_k|h_k|^2 \!\leq \eta \leq \!\bar{P}_{k+1}|h_{k+1}|^2\!\}, ~\forall k\in \{0\}\cup\cal K,
\end{align}
where we define $\bar{P}_{0}|h_{0}|^2\triangleq0$ and $\bar{P}_{K+1}|h_{K+1}|^2\to\infty$ for notational convenience.
Then, it is easy to establish the equivalence between the following two sets:
\begin{align}\label{static_eta_spec_domian}
\{\eta\geq 0\}=\bigcup \limits_{k\in \{0\}\cup\cal K} {\cal F}_k.
\end{align}
Given \eqref{static_eta_spec_domian}, we note that solving problem \eqref{static_K_eta} is equivalent to solving the following $K+1$ subproblems and comparing their optimal values to obtain the minimum one:
 \begin{align}\label{F_k}
\min_{ \eta\in{\cal F}_k} F_k(\eta)\! \triangleq \!\sum_{i=1}^k\! \left(\!\frac{\sqrt{\bar{P}_i}|h_{i}|}{\sqrt{\eta}}\!-\!1\!\!\right)^2\!+\!\frac{\sigma^2}{\eta}, ~\forall k\in\{0\}\cup\cal K,\!
\end{align}
with $F_k(\eta)$ denoting the objective function of the $k$-th subproblem. Notice that when $k =0$, we have $F_0(\eta) =  \sigma^2/\eta$. In addition, we have
\begin{align}\label{static_F_Fk}
	F(\eta) = F_k(\eta), ~\forall \eta \in {\cal F}_k,~&\forall k\in\{0\}\cup\cal K.
\end{align}
More specifically, after solving each subproblem in \eqref{F_k}, we can obtain the optimal solution to problem \eqref{static_K_eta} by comparing their optimal values $\{F_k(\eta_k^*)\}$ in \eqref{F_k}, where $\eta_k^* \in {\cal F}_k$ denotes the optimal denoising factor in each interval $k\in\{0\}\cup\cal K$.
Let $\eta^*$ denote the globally optimal solution for problem \eqref{static_K_eta}. First, we have the following lemma.
\begin{lemma} \label{opt_noactive_lemma}\vspace{-0.1cm}\emph{
Given \eqref{static_order}, the optimal denoising factor must satisfy $\eta^*\geq\bar{P}_1|h_1|^2$, such that device $1$ (that with the smallest quality indicator) should always transmit with full power, i.e., $p_1^*=\bar{P}_1$.}
\end{lemma}\vspace{-0.3cm}
\begin{IEEEproof}
	See Appendix~\ref{opt_noactive_proof}.
\end{IEEEproof}
Lemma~\ref{opt_noactive_lemma} suggests that
we only need to focus on solving each $k$-th subproblem in \eqref{F_k} with $k \in \cal K$, for which we have the following lemma.
\begin{lemma}\vspace{-0.1cm}\label{de_in_function_lemma}\emph{
	For any $k\in \cal K$, the function $F_k(\eta)$ is a unimodal function that first decreases in $[0, \tilde \eta_{k}]$ and then increases in $[\tilde \eta_{k}, \infty)$,
where $\tilde \eta_{k}$ is the stationary point given by
	 \begin{align*}
	\tilde \eta_{k}&=\left( \frac{\sigma^2+ \sum_{i=1}^{k} \bar{P}_{i}|h_{i}|^2}{\sum_{i=1}^{k}\sqrt{\bar{P}_{i}}|h_{i}|}\right)^2.
\end{align*}
Therefore, the optimal solution to the $k$-th subproblem in \eqref{F_k} is given by
\begin{align}\label{static_eta_k}
	\eta_k^*=\min\left( \bar{P}_{k+1}|h_{k+1}|^2, ~\max\left( \tilde \eta_{k},~\bar{P}_k|h_k|^2 \right)\right).
\end{align}
	 }
\end{lemma}
\begin{IEEEproof}
	See Appendix~\ref{de_in_function_proof}.
\end{IEEEproof}
By comparing the optimal values $\{F_k(\eta_k^*)\}$ of the $K$ subproblems in \eqref{F_k} obtained in Lemma~\ref{de_in_function_lemma}, we can obtain the optimal solution to problem \eqref{static_K_eta}. Suppose that $\eta^* = \eta^*_{k^*}\in {\cal F}_{k^*}$, where
\begin{align}\label{static_opt_k_star_F}
	k^*=\arg \min_{k\in\cal K} F_{k}(\eta_k^*).
\end{align}
Accordingly, the optimal solution to problem (P2) is derived as follows.
\begin{theorem} \label{eta_K_static}\vspace{-0.1cm}\emph{
With $k^*$ defined in \eqref{static_opt_k_star_F}, the optimal power control over static channels that solves problem (P2) has a threshold-based structure, given by
\begin{align}
	p_k^*=
	\begin{cases}
		\bar{P}_k,~&\forall k\in\{1,\cdots,k^* \},\\
		\frac{\eta^*}{|h_{k}|^2},~&\forall k\in\{k^*+1,\cdots,K \},
	\end{cases}\label{static_p_lemma}
	\end{align}
where the threshold is given as 	
\begin{align}	
	\eta^*=\tilde \eta_{k^*}=\left( \frac{\sigma^2+ \sum_{i=1}^{k^*} \bar{P}_{i}|h_{i}|^2}{\sum_{i=1}^{k^*}\sqrt{\bar{P}_{i}}|h_{i}|}\right)^2.\label{opt_eta_static}
\end{align}
Furthermore, it holds that $\bar{P}_k|h_k|^2\leq\eta^* $ for devices $k\in\{1,\cdots,k^* \}$ and $\bar{P}_k|h_k|^2\geq\eta^*$ for devices $k\in\{k^*+1,\cdots,K \}$.
}
\end{theorem}
\begin{IEEEproof}
	See Appendix~\ref{eta_K_static_proof}.
\end{IEEEproof}

\begin{remark}[Threshold-Based Optimal Power Control]\label{Threshold_P2}
\emph{
Based on the optimal solution to problem (P2), we have the following insights on the optimal power control for AirComp over devices in static channels.
The optimal power-control policy over devices has a \emph{threshold-based structure}. The threshold is specified by the denoising factor $\eta^*$ and applied on the derived quality indicator $\bar{P}_{k}|h_{k}|^2$ accounting for both the channel power gain and power budget of device $ k\in\cal K$.
It is shown that for each device $k\in \{k^*+1,..., K\}$ with its quality indicator exceeding the threshold, i.e., $\bar{P}_k|h_k|^2\geq\eta^*$, the \emph{channel-inversion power control} is applied with $p_k^* = \frac{\eta^*}{|h_{k}|^2}$; while for each device $k \in \{1, ..., k^*\}$ with $\bar{P}_k|h_k|^2\leq\eta^*$, the \emph{full power transmission} is deployed with $p_k^*=\bar P_k$.
As shown in \eqref{opt_eta_static}, the optimal threshold $\eta^*$ is a monotonically increasing function w.r.t. the noise variance $\sigma^2$. This says that a larger $\sigma^2$ leads to more devices transmitting with full power and vice versa. The result is intuitive by noting that $\eta^*$ also plays another role as the denoising factor: a large $\sigma^2$ requires a large $\eta^*$ for suppressing the dominant noise-induced error.
}
\end{remark}

\vspace{-0.3cm}
\subsection{Alternative Solution Method}
	Though we have found the optimal solution above, in this subsection we further provide an in-depth analysis of $k^*$ to reveal more insights that can be exploited for developing faster algorithms to search for $k^*$.

\begin{lemma}\label{opt_eta_static_ordering_lemma}\vspace{-0.1cm} \emph{
 $k^*$ defined in \eqref{static_opt_k_star_F} has the following properties:
\begin{itemize}
    	\item[1)] $ \bar{P}_{k^*}|h_{k^*}|^2\leq\tilde \eta_{k^*}\leq\bar{P}_{k^*+1}|h_{k^*+1}|^2$;
	\item[2)] $ \bar P_k|h_k|^2\leq\tilde \eta_{k-1},~ \forall k\in\{1,\cdots,k^*\}$ and
	   $\bar P_k|h_k|^2\geq\tilde \eta_{k-1},~\forall k\in\{k^*+1,\cdots,K\}$;
    \item[3)] $k^*=\arg \min_{k\in \cal K} \tilde \eta_{k}$;
\end{itemize}
where we define $\tilde \eta_{0}\to\infty$.
}
\end{lemma}\vspace{-0.3cm}
\begin{IEEEproof}
	See Appendix~\ref{opt_eta_static_ordering_proof}.
\end{IEEEproof}
\begin{figure*}[htbp]
  \centering
    \subfigure[The quality indicator of each device.]{\label{Fig:Power_V_S_ther}\includegraphics[width=8cm]{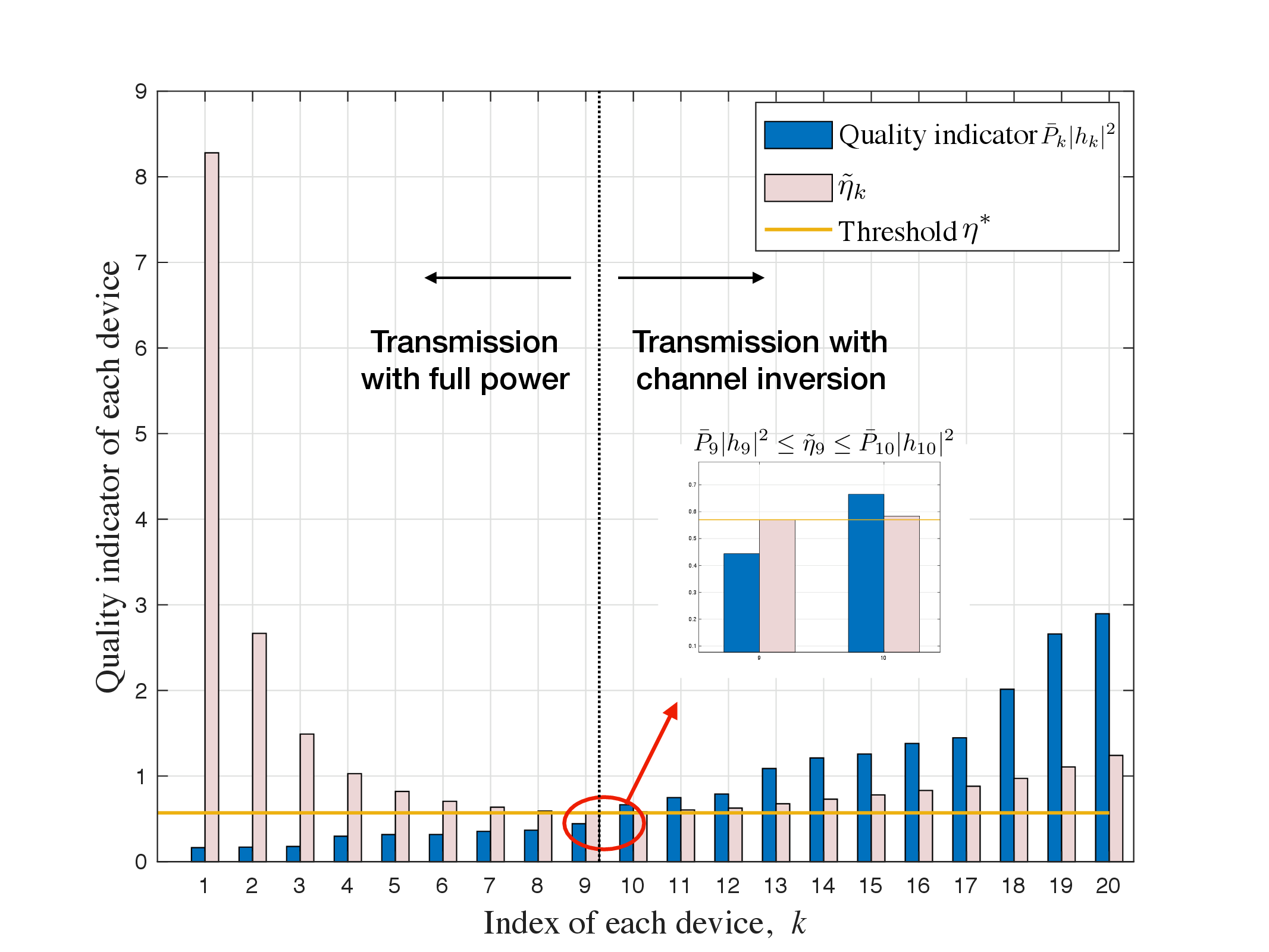}}
  \subfigure[Threshold-based power control among devices.]{\label{fig:Power_V_S_P}
\includegraphics[width=8cm]{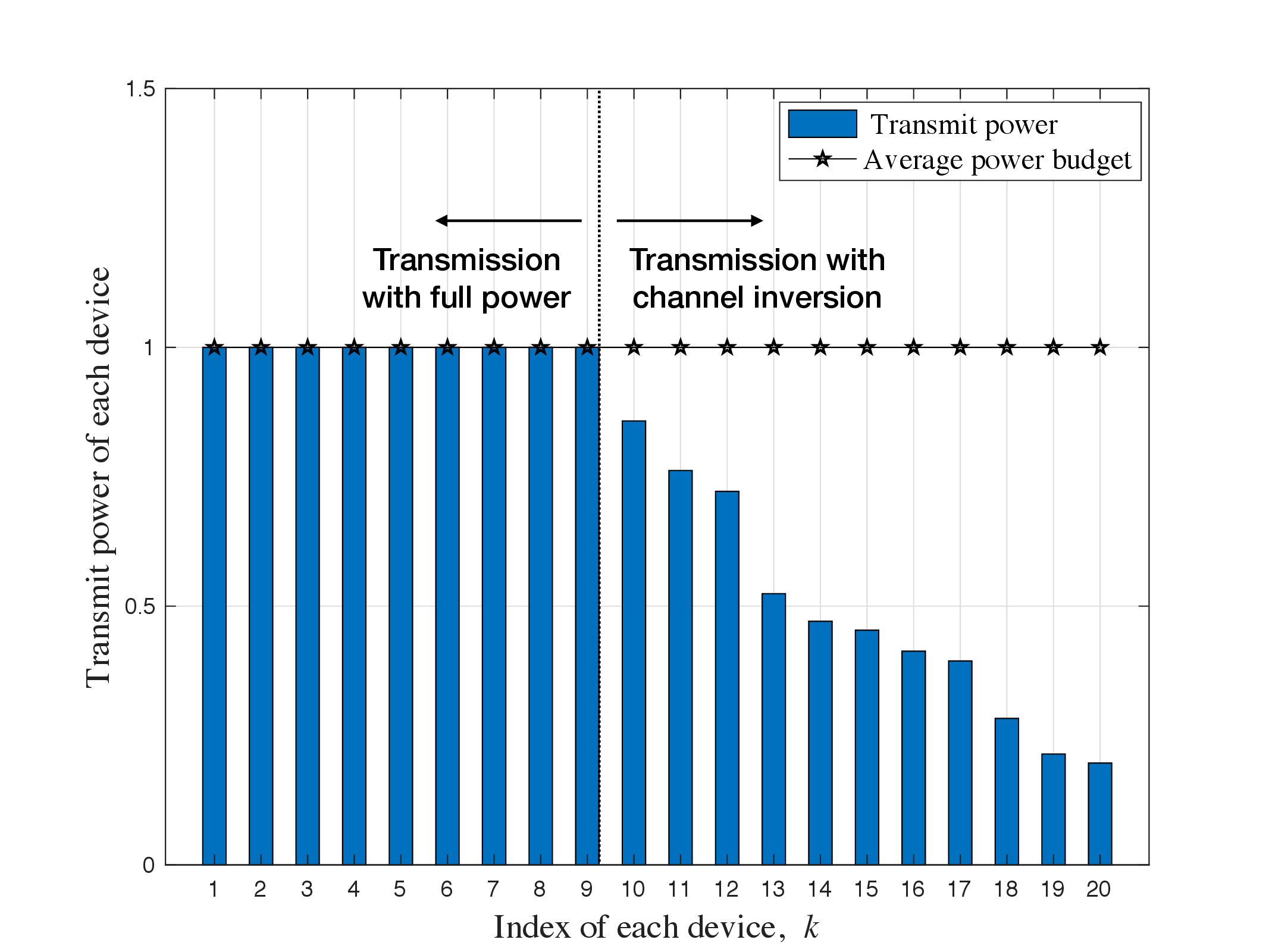}}
  \caption{Optimal power control in static channels with $K=20$ and $\bar P_k = 1$ W, $\forall k\in\cal K$, where the device indices are arranged to meet the ordering assumption in \eqref{static_order}. }
  \label{Fig:fading_P}
\vspace{-0.5cm}
\end{figure*}

The three properties of $k^*$ shown in Lemma~\ref{opt_eta_static_ordering_lemma} suggest three different ways to search for a unique $k^*$. Each property involves simple computation of $\{\bar P_k|h_k|^2\}$ and $\{\tilde \eta_{k}\}$ without the need to further solve the subproblems in \eqref{F_k} to find $\{F_k(\eta)\}$, which significantly reduces the complexity of solving problem (P2). Notice that the fastest one may be the one using property $3)$ that only involves computing $\{\tilde \eta_{k}\}$ sequentially and finding the minimum. Once $k^*$ is found, we can directly obtain the optimal transmission power $\{p_k^*\}$ and denoising factor $\eta^*$ by applying Theorem~\ref{eta_K_static}.

The correctness of the derived results in Theorem~\ref{eta_K_static} and Lemma \ref{opt_eta_static_ordering_lemma} is further verified by an experiment under typical system settings plotted in Fig.~\ref{Fig:fading_P}.
Specifically, it is observed from Fig. \ref{Fig:Power_V_S_ther} that $k^*= 9$, as it follows $\bar{P}_{9}|h_{9}|^2\leq\tilde \eta_{9}\leq\bar{P}_{10}|h_{10}|^2$ with $\eta^*=\tilde \eta_{9}=\arg \min_{k\in \cal K} \tilde \eta_{k}$.
Furthermore, for any device $k\in\{1,\cdots,9\}$, we have $\bar P_k|h_k|^2\leq\tilde \eta_{k-1}$; while for any device $k\in\{10,\cdots,20\}$, it holds that $\bar P_k|h_k|^2\geq\tilde \eta_{k-1}$. These observations are consistent with the three properties obtained in Lemma \ref{opt_eta_static_ordering_lemma}.
Besides, as shown in Fig.~\ref{fig:Power_V_S_P}, devices $1-9$ transmit with full power while the others transmit with channel-inversion power control, which confirms the results derived in Theorem \ref{eta_K_static}.
\vspace{-0.1cm}
\subsection{Asymptotic Analysis}\label{static_asym}
In this subsection, we analyze the derived optimal power-control policy to problem (P2) in two extreme regimes, namely the high and low SNR regimes, respectively.

 In the high SNR regime, i.e., $\sigma^2 \rightarrow0$, it is easy to know from problem \eqref{static_K_eta} that the MSE is dominated by the misalignment error. Thus we can expect that the channel-inversion power control is optimal as it minimizes the misalignment error.
Aligned with the intuition, it follows from Lemma \ref{opt_noactive_lemma} that $\eta^* = \bar P_1 |h_1|^2$ should be the global minimizer solving problem \eqref{static_K_eta},
  and correspondingly we have $k^*=1$ in Theorem \ref{eta_K_static}.
Therefore, the optimal power control and denoising factor are given by
\begin{align}\label{opt_two_high}
p_k^*=\frac{\eta^*}{|h_k|^2}, \forall k\in{\cal K},~ {\rm and} ~\eta^* = \bar{P}_1{|h_1|^2}.
\end{align}

In the low SNR regime with $\sigma^2  \rightarrow\infty$, the noise-induced error becomes dominant, and  thereby it is desirable to maximize $\eta$ to suppress the noise-induced error, i.e., $\eta^*\geq \bar{P}_K|h_K|^2$ must hold. Then we have $k^*=K$ in Theorem \ref{eta_K_static}. In this case, all devices should transmit with full power to achieve the optimal $\eta^*$.
 The optimal power control and denoising factor in this case are
\begin{align}\label{opt_two_low}
p_k^*=\bar{P}_k, \forall k\in{\cal K}, ~{\rm and} ~\eta^*=\tilde \eta_{K}.
 \end{align}

 In summary, the above analysis suggests that the optimal power control reduces to the channel inversion and full power transmission in the high and low SNR regimes, respectively. The analytical findings will be substantiated by simulation results in Section \ref{sec_simu}.

\vspace{-0.3cm}
\section{Power Control with Time-Varying Channels}\label{sec_fading}
In this section, we present the power-control policy from solving problem (P1) with time-varying channels. In the following, we first consider the general case with any power budget values of $\{\bar P_k\}$, and then consider a special case with only one device being power limited (i.e., $\bar P_k\!\to\!\infty, \forall k\in \!{\cal K}\!\setminus\! \{1\}$) for gaining useful design insights.
\vspace{-0.5cm}
\subsection{Power Control Policy over Devices and Fading States}
Despite the non-convexity of problem (P1), it can be shown to satisfy the so-called ``time sharing" condition, which refers to a particular condition that when a non-convex problem satisfies, the strong duality holds. The condition holds for many practical communication system where a common finite radio resource is shared over time, giving the name ``time sharing".
For more details, please refer to Definition 1 in \cite{yuwei06}.
To proceed with, strong duality holds between problem (P1) and its Lagrange dual problem \cite{Boyd2004}, and thus we leverage the Lagrange-duality method to optimally solve problem (P1).
Let $\mu_k\ge0$ denote the dual variable associated with the $k$-th constraint in \eqref{fad_bar_P_ave}, $\forall k\in \cal K$.
Then the partial Lagrangian of problem (P1) is
\begin{align*}
&\mathcal{L}(\!\{ p_k(\bm\nu),\eta(\bm\nu), \mu_k\} )\\
&\!\!=\!\mathbb{E}_{\bm\nu}\!\!\left[\!\sum \limits_{k\in{\cal K}}\!\!\left(\!\frac{\sqrt{p_{k}(\!\bm\nu\!)}|h_{k}|}{\sqrt{\eta(\bm\nu)}}\!-\!1\!\right)^2\!\!\!\!+\!\frac{\sigma^2}{\eta(\!\bm\nu\!)}\!\right]\!\!+\! \sum \limits_{k\in\cal K}\!\mu_k\!\left(\!\mathbb{E}_{\bm\nu}\![p_k(\bm\nu)]\!-\!\bar{P}_k\!\right)\!.
\end{align*}
The dual function is
\begin{align}
G(\{ \mu_k\} )=\min_{\{ p_k(\bm\nu)\ge 0\},\{\eta(\bm\nu)\ge 0\}}& ~\mathcal{L}(\{ p_k(\bm\nu),\eta(\bm\nu), \mu_k\} ).\label{fading_dualfunction}
\end{align}
It is observed that the dual function in \eqref{fading_dualfunction} is linear w.r.t. the dual variable $ \mu_k$ but non-convex w.r.t. $p_k(\bm\nu)$ and $\eta(\bm\nu)$.
Accordingly, the dual problem of problem (P1) is given as
\begin{align}
\mathbf{(D1):} \max_{\{ \mu_k\geq0\}}& ~G(\{ \mu_k\} ).
\end{align}
Since the strong duality holds between problems (P1) and (D1), one can solve problem (P1) by equivalently solving its dual problem (D1) \cite{Boyd2004}.
For notational convenience, let  $\{ p_k^{\rm opt}(\bm\nu)\}$ and $\{\eta^{\rm opt}(\bm\nu)\}$ denote the optimal primal solution to problem (P1), and $\{ \mu_k^{\rm opt}\}$ denote the optimal dual solution to problem (D1).
In the following, we first evaluate the dual function $G(\{ \mu_k\} )$ with any given feasible $\{ \mu_k\}$, and then obtain the optimal dual solution $\{ \mu_k^{\rm opt}\}$ to maximize $G(\{ \mu_k\} )$.


First, we obtain $G(\{ \mu_k\} )$ by solving problem \eqref{fading_dualfunction} with any given $\{ \mu_k\ge 0\}$, which can be decomposed into a sequence of subproblems each for one particular  fading state $\bm\nu$ as follows:
\begin{align}\label{fading_sub_dualfunction}
\min_{\substack{\{p_k(\bm\nu)\geq0\},\\
	  \eta(\bm\nu)\geq0}}~& \!\!\sum \limits_{k\in\cal K}\!\!\left( \!\frac{\sqrt{p_{k}(\bm\nu)}|h_{k}|}{\sqrt{\eta(\bm\nu)}}\! -\! 1\!\!\right)^2\!\!\!+\!\! \frac{\sigma^2}{\eta(\bm\nu)}\!\!+\!\!\sum \limits_{k\in\cal K}\!\mu_kp_k(\bm\nu).
\end{align}
Note that problem \eqref{fading_sub_dualfunction} is non-convex. To solve this problem, we first consider the optimization over $\{p_k(\bm\nu)\geq0\}$ with given $\eta(\bm\nu)$, which can be decomposed into $K$ subproblems as follows, each for one device $k\in \cal K$:
\begin{align}\label{fading_sub_over_K}
\min_{p_k(\bm\nu)\geq0}& ~\left(\frac{\sqrt{p_{k}(\bm\nu)}|h_{k}|}{\sqrt{\eta(\bm\nu)}}-1\right)^2+\mu_kp_k(\bm\nu).
\end{align}
\begin{lemma}\label{opt_p_lemma}\emph{
	The optimal solution to problem \eqref{fading_sub_over_K} is given as
\begin{align}\label{opt_Pkv}
p_k^*(\bm\nu)=\left(\frac{\sqrt{|h_{k}|^2\eta(\bm\nu)}}{|h_{k}|^2+\eta(\bm\nu)\mu_k} \right)^2.
\end{align}}
\end{lemma}
By replacing $\{p_k(\bm\nu)\}$ with $\{p_k^*(\bm\nu)\}$ in Lemma \ref{opt_p_lemma}, problem \eqref{fading_sub_dualfunction} becomes a single-variable optimization problem over $\eta(\bm\nu)$:
\begin{align}\label{fading_sub_over_eta1}
\min_{\eta(\bm\nu)\geq0}& ~\sum \limits_{k\in\cal K}  \frac{\eta(\bm\nu)\mu_k }{|h_{k}|^2+\eta(\bm\nu)\mu_k}+\frac{\sigma^2}{\eta(\bm\nu)}.
\end{align}
Let $\gamma(\bm\nu)=1/{\eta(\bm\nu)}$, and thus problem \eqref{fading_sub_over_eta1} becomes a convex optimization problem over $\gamma(\bm\nu)$, which is shown as
\begin{align}\label{fading_sub_over_gamma}
	\min_{\gamma(\bm\nu)\geq0}& ~\sum \limits_{k\in\cal K} \frac{1}{ \lambda_k(\bm\nu)\gamma(\bm\nu)+1}+\gamma(\bm\nu){\sigma^2},
\end{align}
with $\lambda_k(\bm\nu) \triangleq |h_{k}|^2/\mu_k$.
By checking the first-order derivative of the objective function in problem \eqref{fading_sub_over_gamma} w.r.t. $\gamma(\bm\nu)$, it follows that the optimal solution $\gamma^*(\bm\nu)$ to problem \eqref{fading_sub_over_gamma} must satisfy the following equality:
\begin{align}\label{fading_sub_over_gamma_firstorderderi}
\sum \limits_{k\in\cal K}\frac{\lambda_k(\bm\nu)}{\left(\lambda_k(\bm\nu)\gamma^*(\bm\nu)+1\right)^2}=\sigma^2.
\end{align}
Based on \eqref{fading_sub_over_gamma_firstorderderi}, the optimal $\gamma^*(\bm\nu)$ can be efficiently found by a bisection search, though a closed-expression is not admitted. As a result, the optimal solution to problem \eqref{fading_sub_over_eta1} is obtained as $\eta^*(\bm\nu)=1/\gamma^*(\bm\nu)$.
With the optimal $\eta^*(\bm\nu)$ in hands, we can get the optimal power control $\{p_k^*(\bm\nu)\}$ based on \eqref{opt_Pkv} in Lemma \ref{opt_p_lemma}, by replacing $\eta(\bm\nu)$ as $\eta^*(\bm\nu)$.
Therefore, problem \eqref{fading_sub_dualfunction} is finally solved with any given $\{ \mu_k\} $, and the dual function $G(\{ \mu_k\} )$ is also accordingly obtained.


Next, we search over $\{ \mu_k\ge 0\}$ to maximize $G(\{ \mu_k\} )$ for solving problem (D1). Since the dual function $G(\{ \mu_k\} )$ is concave but non-differentiable in general, one can use subgradient based methods such as the ellipsoid method \cite{ellipsoid}, to obtain the optimal $\{ \mu_k^{\rm opt}\}$ for problem (D1). Notice that for the objective function in problem \eqref{fading_dualfunction}, the subgradient w.r.t. $\mu_k$ is $\mathbb{E}_{\bm\nu}(p_k^*(\bm\nu))-\bar{P}_k$, $\forall k\in\cal K$.
By replacing $\{\mu_k\}$ in Lemma \ref{opt_p_lemma} as the obtained optimal $\{ \mu_k^{\rm opt}\}$, the optimal solution to (P1) is presented as follows.\vspace{-0.1cm}
\begin{theorem}\label{P1_opt_sol}\emph{
	The optimal power-control policy over devices and fading states to problem (P1) are given as
\begin{align*}
	p_k^{\rm opt}(\bm\nu)=\left(\frac{\sqrt{|h_{k}|^2\eta^{\rm opt}(\bm\nu)}}{|h_{k}|^2+\eta^{\rm opt}(\bm\nu)\mu_k^{\rm opt}} \right)^2,~\forall k\in{\cal K}, ~\forall \bm\nu,
\end{align*}
	where $\{\eta^{\rm opt}(\bm\nu)\}$ can be obtained via a bisection search based on \eqref{fading_sub_over_gamma_firstorderderi} with $\{ \mu_k^{\rm opt}\}$. }
\end{theorem}
\begin{remark}[Regularized Channel Inversion Power Control]\vspace{-0.1cm}\emph{
The optimal power control $\{p_k^{\rm opt}(\bm\nu)\}$ in Theorem \ref{P1_opt_sol} are observed to follow an interesting regularized channel inversion structure, with denoising factors $\{\eta^{\rm opt}(\bm\nu)\}$ and dual variables $\{ \mu_k^{\rm opt}\}$ acting as parameters for regularization.
More specifically, it is observed that for any device $k\in {\cal K}$, if $\mu_k^{\rm opt}>0$ holds, the average power constraint of device $k$ must be tight at the optimality due to the complementary slackness condition (i.e., $\mu_k^{\rm opt}(\mathbb{E}_{\bm\nu}[p_k^{\rm opt}(\bm\nu)]-\bar{P}_k)=0$), and thus this device should use up its transmit power budget based on the regularized channel inversion power control over fading states; otherwise, with $\mu_k^{\rm opt} = 0$, device $k$ should transmit with channel-inversion power control without using up its power budget.
}
\end{remark}
Furthermore, the following theorem shows that with Rayleigh fading channels, the average power constraints at all devices should be tight. Intuitively, this is due to the fact that in this case, deep channel fading may occur over time, and thus sufficiently large transmit power is required for implementing the regularized channel inversion power control.
\begin{theorem}[Rayleigh Fading]\vspace{-0.1cm}\label{general_K_activenum_proof}\emph{
	With Rayleigh fading, where the channel coefficient $h_k$'s are independent \emph{circularly symmetric complex Gaussian} (CSCG) random variables with zero mean and variance of $\sigma_h^2$, i.e., $h_k\sim{\cal CN}(0,\sigma_h^2)$, $\forall k\in \cal K$, it must hold at the optimal solution to problem (P1) that $\mathbb{E}_{\bm\nu}[p_k^{\rm opt}(\bm\nu)]=\bar{P}_k, \forall k\in \cal K$. In other words, the average power constraints must be tight for all the $K$ devices.}
\end{theorem}
\vspace{-0.1cm}
\begin{IEEEproof}\vspace{-0.1cm}
See Appendix~\ref{general_K_activenum}.
\end{IEEEproof}

\vspace{-0.2cm}
\subsection{Special Case with Only One Power-Limited Device}
\vspace{-0.1cm}
To get more insights, we consider a special $K$-device case, in which only one device, say device $1$, is subject to the average power constraint and all the other devices have sufficiently high power budgets (i.e., $\bar P_k \to \infty, \forall k\in {\cal K}\setminus \{1\}$).
Note that such a special case is of practical relevance.
One example is that one device has low battery power compared with others.
As another example, there may exist one device located in the remote region farther away the FC than other devices.
In such cases, other devices need to reduce their power to meet the signal-magnitude alignment requirement, and the said device becomes the bottleneck of the system (more severe path loss). As a result, the transmit power constraints at all the other devices may become inactive at the optimal point.
This is thus equivalent to the case with one single transmit power constraint.
To proceed with, problem (P1) reduces to
\begin{align}
\mathbf{(P3):}\!\min_{\{ p_k(\bm\nu)\ge 0\}\!,\!\{\eta(\bm\nu)\ge 0\}\!} & \mathbb{E}_{\bm\nu}\!\left[\!\sum \limits_{k\in{\cal K}}\!\!\left(\!\frac{\sqrt{p_{k}(\bm\nu)}|h_{k}|}{\sqrt{\eta(\bm\nu)}}-\!1\!\right)^2\!\!+\!\frac{\sigma^2}{\eta(\bm\nu)}\!\right]\notag\\
{\rm s.t.}~~~~~~~&\mathbb{E}_{\bm\nu}[p_1(\bm\nu)]\leq \bar{P}_1\label{P3_power}.
\end{align}
First, as each device $ k\in {\cal K}\setminus \{1\}$ has unconstrained transmit power, it is evident that with any given $\{\eta(\bm\nu)\}$, the optimal power control has the following channel inversion structure:
\begin{align}\label{two_case_SubP2_opt_Pkn}
p_{k}^{\star} (\bm\nu)=\frac{\eta(\bm\nu)}{|h_{k}|^2},~ \forall k\in {\cal K} \setminus \{1\}, ~\forall \bm\nu.
\end{align}
By substituting the optimal $p_k^{\star}(\bm\nu), \forall k\in {\cal K}\setminus \{1\}$, $\forall \bm\nu$, problem (P3) boils down to the optimization over $\eta(\bm\nu)$'s and $p_1(\bm\nu)$'s, i.e.,
\begin{align}\label{K2_sub_P1etaunderP2}
\min_{\{ p_1(\bm\nu)\geq0\},\{\eta(\bm\nu)\geq0\}} ~& \!\mathbb{E}_{\bm\nu} \!\left[\left(\frac{\sqrt{p_{1}(\bm\nu)}|h_{1}|}{\sqrt{\eta(\bm\nu)}} \!- \!1\right)^2 \!\!+ \!\frac{\sigma^2}{\eta(\bm\nu)}\right] \\
{\rm s.t.}~~~~~~~~&\mathbb{E}_{\bm\nu}[p_1(\bm\nu)]\leq \bar{P}_1. \notag
\end{align}

Next, for problem \eqref{K2_sub_P1etaunderP2}, we first optimize $\{\eta(\bm\nu)\}$ with any given $\{p_1(\bm\nu)\}$, in which the optimal $\{\eta(\bm\nu)\}$ is given as
\begin{align}
\eta^{\star}(\bm\nu)=\left( \frac{\sigma^2+ p_{1}(\bm\nu)|h_{1}|^2}{\sqrt{p_{1}(\bm\nu)}|h_{1}|}\right)^2, \forall \bm\nu.\label{two_case_eta_opt}
\end{align}
By substituting $\{\eta^{\star}(\bm\nu)\}$ into problem \eqref{K2_sub_P1etaunderP2}, the remaining optimization over $\{p_1(\bm\nu)\}$ corresponds to the following convex optimization problem:
\begin{align}\label{two_case_P1_OPT}
\min_{\{ p_1(\bm\nu)\geq0\}} ~& \mathbb{E}_{\bm\nu}\left[\frac{\sigma^2}{\sigma^2+p_{1}(\bm\nu)|h_{1}|^2}\right]\\
{\rm s.t.}~~~&\mathbb{E}_{\bm\nu}[p_1(\bm\nu)]\leq \bar{P}_1.\notag
\end{align}
The optimal solution to problem \eqref{two_case_P1_OPT} is obtained by applying the Karush-Kuhn-Tucker (KKT) conditions, as given in the following.
\begin{lemma}\label{Specific_K_lemma}\vspace{-0.3cm}\emph{
The optimal power-control policy of the power-limited device $1$ in problem \eqref{two_case_P1_OPT} is given by
\begin{align}
	p_{1}^{\rm opt}(\bm\nu)&=\frac{\sqrt{\sigma^2} }{\sqrt{ \mu_1^{\rm opt}} |h_1|^2}\left(|h_1|- \sqrt{ \mu_1^{\rm opt}\sigma^2} \right)^+, ~\forall \bm\nu,  \label{K2_opt_p_1n}
\end{align}
where $(x)^+=\max\{0,x\}$ and $\mu_1^{\rm opt}$ denotes the optimal Lagrange multiplier associated with the average power constraint of \eqref{P3_power} that can be obtained via a bisection search to enforce the equality $\mathbb{E}_{\bm\nu}(p_{1}^{\rm opt}(\bm\nu)) = \bar P  _1$.}
\end{lemma}
\begin{IEEEproof}
See Appendix~\ref{Specific_K_proof}.
\end{IEEEproof}
Combining \eqref{two_case_SubP2_opt_Pkn}, \eqref{two_case_eta_opt}, and Lemma~\ref{Specific_K_lemma}, the optimal solution to problem (P3) is summarized in the following theorem.
\begin{theorem}\label{P3_lem}\vspace{-0.1cm}\emph{
For the special case with device $1$ being power limited, the optimal power-control policy to problem (P3) are given as
\begin{align*}
p_{1}^{\rm opt}(\bm\nu)&=\frac{\sqrt{\sigma^2} }{\sqrt{ \mu_1^{\rm opt}} |h_1|^2}\left(|h_1|- \sqrt{ \mu_1^{\rm opt}\sigma^2} \right)^+, ~\forall \bm\nu,\\
p_{k}^{\rm opt}(\bm\nu)&=\frac{\eta^{\rm opt}(\bm\nu)}{|h_k|^2},~\forall k\in{\cal K}\setminus \{1\}, ~\forall \bm\nu.
\end{align*}
The optimal denoising factors are
\begin{align*}
	\eta^{\rm opt}(\bm\nu)&=\left( \frac{\sigma^2+ p_{1}^{\rm opt}(\bm\nu)|h_{1}|^2}{\sqrt{p_{1}^{\rm opt}(\bm\nu)}|h_{1}|}\right)^2,~ \forall \bm\nu.
\end{align*}
}
\end{theorem}\vspace{-0.3cm}
From Theorem 4, it is first evident that we have $p_1^{\rm opt}(\bm\nu) = 0$ and $\eta^{\rm opt}(\bm\nu) \to \infty$ when the channel gain of device $1$ is no larger than the threshold $\sqrt{\mu_1^{\rm opt}\sigma^2}$, i.e., $|h_1|\le\sqrt{\mu_1^{\rm opt}\sigma^2}$.
Furthermore, by taking the first-order derivative of $p_{1}^{\rm opt}(\bm\nu)$ w.r.t. $|h_1|$,  it is observed that the optimal power control $p_{1}^{\rm opt}(\bm\nu)$ is first increasing in $|h_1| \in\left(\sqrt{\mu_1^{\rm opt}\sigma^2},2\sqrt{\mu_1^{\rm opt}\sigma^2} \right]$ and then decreasing in $|h_1| \in\left[2\sqrt{\mu_1^{\rm opt}\sigma^2},\infty \right)$.
Substituting $p_{1}^{\rm opt}(\bm\nu)$ into $\eta^{\rm opt}(\bm\nu)$, it is also observed that $\eta^{\rm opt}(\bm\nu)$ is first decreasing in $|h_1| \in\left(\sqrt{\mu_1^{\rm opt}\sigma^2},2\sqrt{\mu_1^{\rm opt}\sigma^2} \right]$ and then increasing in $|h_1| \in\left[2\sqrt{\mu_1^{\rm opt}\sigma^2},\infty \right)$.
\begin{figure}
\centering
 \setlength{\abovecaptionskip}{-1mm}
\setlength{\belowcaptionskip}{-1mm}
    \includegraphics[width=8cm]{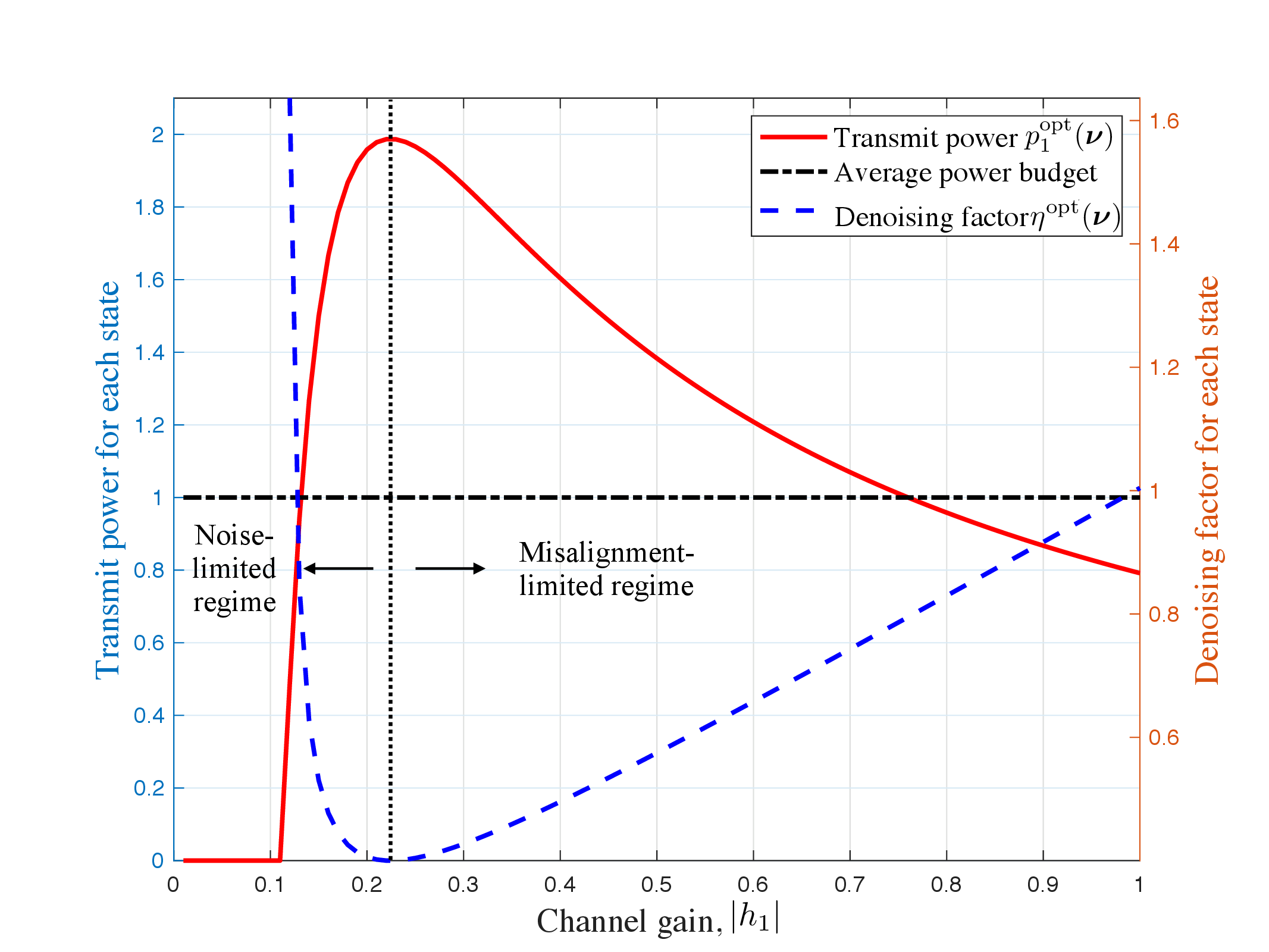}
\caption{Optimal power control of power-limited device $1$ with $\bar P_1=1$ W and $\sigma^2=1$, where the wireless channels follow  i.i.d. Rayleigh fading with average channel gains normalized.} \label{fig:Power_fading}
\vspace{-0.1cm}
\end{figure}
\begin{remark}[Channel-Inversion Water-Filling Power Control]\vspace{-0.1cm}\emph{
It is observed from Theorem~\ref{P3_lem} that the optimal power control of the power-limited device $1$ follows a water-filling type structure: there exists a water level (i.e., $\sqrt{\mu_1^{\rm opt}\sigma^2}$), such that when the channel gain is larger than this level, it transmits with positive power; otherwise, it keeps silent without transmission. Furthermore, once transmitting, the transmit power is inversely proportional to the channel gain. Therefore, we name this solution as \emph{channel-inversion water-filling} power control. The specialty of the solution lies in the non-monotonic relationship between the transmit power and the channel gain. Specifically, as the channel gain decreases till $2\sqrt{\mu_1^{\rm opt}\sigma^2}$, the power-limited device $1$ should increase the transmit power $p_1^{\rm opt}(\bm\nu)$ for the purpose of compensating the channel fading to enforce signal-magnitude alignment. When the channel gain continues to decrease from $2\sqrt{\mu_1^{\rm opt}\sigma^2}$ to $\sqrt{\mu_1^{\rm opt}\sigma^2}$, the power-limited device $1$ should reduce $p_1^{\rm opt}(\bm\nu)$ as the channel is too noisy and thus costly to be compensated. When the channel gain is smaller than $\sqrt{\mu_1^{\rm opt}\sigma^2}$, device $1$ keeps silent to abandon the channel in deep fade to save energy for other fading states. This is in sharp contrast to the conventional water-filling power control for throughput maximization in point-to-point fading channels, with monotonically non-decreasing power control w.r.t. the channel gains.
The derived optimal structure is verified by simulation results shown in Fig.~\ref{fig:Power_fading} under typical system settings, where the transmit power of power-limited device $1$ is plotted under varying channel gains (see the solid curve).
}
\end{remark}
\begin{remark}[Misalignment Error versus Noise Enhancement]\vspace{-0.1cm}\emph{
As suggested by Theorem~\ref{P3_lem} and illustrated in Fig.~\ref{fig:Power_fading}, the optimal denoising factor $\{\eta^{\rm opt}(\bm\nu)\}$ exhibits a ``V'' shape w.r.t. the channel gain: as the channel gain decreases, $\eta^{\rm opt}(\bm\nu)$ sees a trend of first decreasing and then increasing.
The phenomenon can be better explained by dividing the channel gain values into two regimes (see Fig.~\ref{fig:Power_fading}), namely, the noise-limited regime (with low channel gains) and the misalignment-limited regime (with high channel gains).
The two regimes put different priorities on suppressing noise and enforcing signal-magnitude alignment for computation error minimization. Specifically, in the noise-limited regime, the FC should increase the denoising factor $\eta^{\rm opt}(\bm\nu)$ for noise suppression, while signal-magnitude alignment becomes secondary. In contrast, in the misalignment-limited regime, the computation error is dominated by the misalignment error. Thus the optimal power control makes best effort to align signal-magnitude but at a cost of elevating noise-induced error. Similarly, the analytical result is confirmed by simulation results shown in Fig.~\ref{fig:Power_fading}, where the denoising factor is plotted against the varying channel gains (see the dashed line).
}
\end{remark}

\vspace{-1cm}
\subsection{Low-Complexity Power Control Design}
\vspace{-1mm}
So far, we have obtained the optimal solution to problem (P1) with fading channels, where both the transmit power at devices and denoising factors at the FC  depend on the instantaneous channel gains. Therefore, the implementation of such an optimal design requires the FC to collect the CSI from all devices and broadcast the adopted denoising factors at each fading state for power control. This, however, leads to frequent signaling with large overhead.
To tackle this issue, in this subsection, we propose a low-complexity design, in which each device implements the truncated-channel-inversion power control over a portion of fading states (with large channel gains) to perfectly align the signal in magnitude, and the truncation threshold $\xi_k$ at each device $k\in\cal K$ is designed independently, such that the device's available transmit power is fully used for reducing the computation error.
Besides, the FC sets the denoising factors unchanged over different fading states, i.e., $\eta(\bm\nu) = \eta, \forall \bm\nu$, to minimize the signaling overhead.
In this case, we have the power control as
\begin{align}
	p_k(\bm\nu)=
	\begin{cases}
		\frac{\eta}{|h_{k}|^2}, & ~{\rm if} ~|h_{k}|^2\geq \xi_k,\\
		0, &~{\rm otherwise}.
			\end{cases}
\end{align}
Accordingly, the MSE in problem (P1) is re-expressed  as
\begin{align}
\!\!\!\mathbb{E}_{\bm\nu}\!\!\!\left[\!\sum \limits_{k\in{\cal K}}\!\!\left(\!\frac{\sqrt{p_{k}(\!\bm\nu\!)}|h_{k}|}{\sqrt{\eta}}\!\!-\!1\!\right)^2\!\right]\!\!\!+\!\frac{\sigma^2}{\eta}\!=\!K\!- \!\!\sum \limits_{k\in\cal K}\!\mathbb{E}_{\bm\nu}\!\left[\!I_k\!(\!\bm\nu\!) \!\right]\!+\!\frac{\sigma^2}{\eta}, \!  
\end{align}
where $I_k(\bm\nu)$ is defined as an indicator function for the event of channel inversion based on $|h_{k}|^2$ at each fading state $\bm\nu$, shown as
\begin{align}
	I_k(\bm\nu)=
	\begin{cases}
		1, &  ~{\rm if} ~|h_{k}|^2\geq \xi_{k},\\
		0, &~{\rm otherwise}.
			\end{cases}
\end{align}
Under the above truncated-channel-inversion power control design, problem (P1) reduces to a joint optimization problem over $\eta$ and $\{\xi_{k}\}$, i.e.,
\begin{align}
\min_{\eta\geq0,\{\xi_{k}\geq0\}} ~& K- \sum \limits_{k\in\cal K}\mathbb{E}_{\bm\nu}(I_k(\bm\nu) )+\frac{\sigma^2}{\eta} \label{P4_1}\\
{\rm s.t.}~~~~&\mathbb{E}_{\bm\nu}\left[\frac{ I_k(\bm\nu)\eta}{|h_{k}|^2}  \right] \leq \bar{P}_k,~\forall k\in\cal K. \label{K_sub_eta_expect}
\end{align}
To solve problem \eqref{P4_1}, we first find $ \{\xi_{k}\}$ with given $\eta$ and then search for the optimal $\eta^*$. Note that at the optimality, the constraints in \eqref{K_sub_eta_expect} must be tight. Thus, we have
\begin{align}
	\mathbb{E}_{\bm\nu}\left[\frac{ I_k(\bm\nu){\eta}}{|h_{k}|^2}  \right] = \bar{P}_k,~\forall k\in\cal K. \label{g_k_fixed_eta}
\end{align}
With any given $\eta$, the optimal $\{\xi_k^*\}$ is obtained by a bisection search based on \eqref{g_k_fixed_eta}. Substituting $\{\xi_k^*\}$ into problem \eqref{P4_1}, we can then obtain $\eta^*$ accordingly by a one-dimensional search.

\begin{remark}[Signaling Overhead Reduction]\vspace{-0.1cm}
\emph{
Note that the optimal power control requires the instantaneous CSI to design the real-time $\eta(\bm\nu)$ and  power control $\{p_k(\bm\nu)\}$ over each fading state against the channel fluctuation in a centralized manner at the FC.
Unlike the optimal power control, the low-complexity design only requires channel distribution information at the FC to design a uniform $\eta$ over all fading states in an {\it offline} manner.
Via broadcasting $\eta$ to devices prior to transmission, $\xi_k$ can be found at each device locally based on \eqref{g_k_fixed_eta} without any real-time signaling. Therefore, this dramatically reduces the implementation complexity and overhead needed for power control, but at a cost of compromised performance.
}
\end{remark}
\vspace{-0.4cm}
\section{Simulation Results}\label{sec_simu}

In this section, we provide simulation results to validate the performance of the proposed power control for AirComp over fading channels.
In the simulation, the wireless channels from each device to the FC over fading states follow i.i.d. Rayleigh fading, such that $h_k$'s are modeled as i.i.d. CSCG random variables with zero mean and unit variance.
 For each device $k\in{\cal K}$, we define $\rho_k \triangleq \mathbb{E}\left[\frac{\bar P_k|h_k|^2}{\sigma^2}\right]= \bar P_k/\sigma^2$ as the expected receive SNR.
 We apply the Monte-Carlo method to obtain numerical results by averaging over $10^4$ random realizations.

\vspace{-0.1cm}
\subsection{Benchmark Schemes}
In the case with static channels, we consider two benchmark schemes for performance comparison, namely the {\it full power transmission} defined in \eqref{opt_two_low}, and the {\it traditional-channel-inversion} modified from \eqref{opt_two_high} by including a truncation operation that cuts off the channel if its gain is below a pre-specified threshold $\xi$. The truncation operation is added to account for the practical fact that a channel in deep fade will cause outage and thus should be abandoned in real system implementation. 

In the case with time-varying channels, we consider the  {\it uniform power control} and {\it  traditional-channel-inversion} as two benchmark schemes for comparison. To be specific, in the uniform power control scheme, each device transmits with fixed power equal to the expected power budget regardless the channel state, namely $p_k(\bm\nu)=\bar{P}_k, \forall \bm\nu$, where a uniform denoising factor in the FC is applied as $\eta=\min_{k\in\cal K} \mathbb{E}_{\bm\nu}[\bar P_k|h_k|^2]$ from Lemma~\ref{opt_noactive_lemma}.
The traditional-channel-inversion in the current case implements its counterpart for static case separately for each fading state with a uniform instantaneous power constraint set to be its expected value $\bar P_k$.

\vspace{-0.2cm}
\subsection{Performance of AirComp in Static Channels}
\vspace{-0.1cm}
\begin{figure}[htbp]\centering
  \subfigure[Uniform expected receive SNR case.]
  {\label{fig:MSE_S_V_K_uni}\includegraphics[width=8cm]{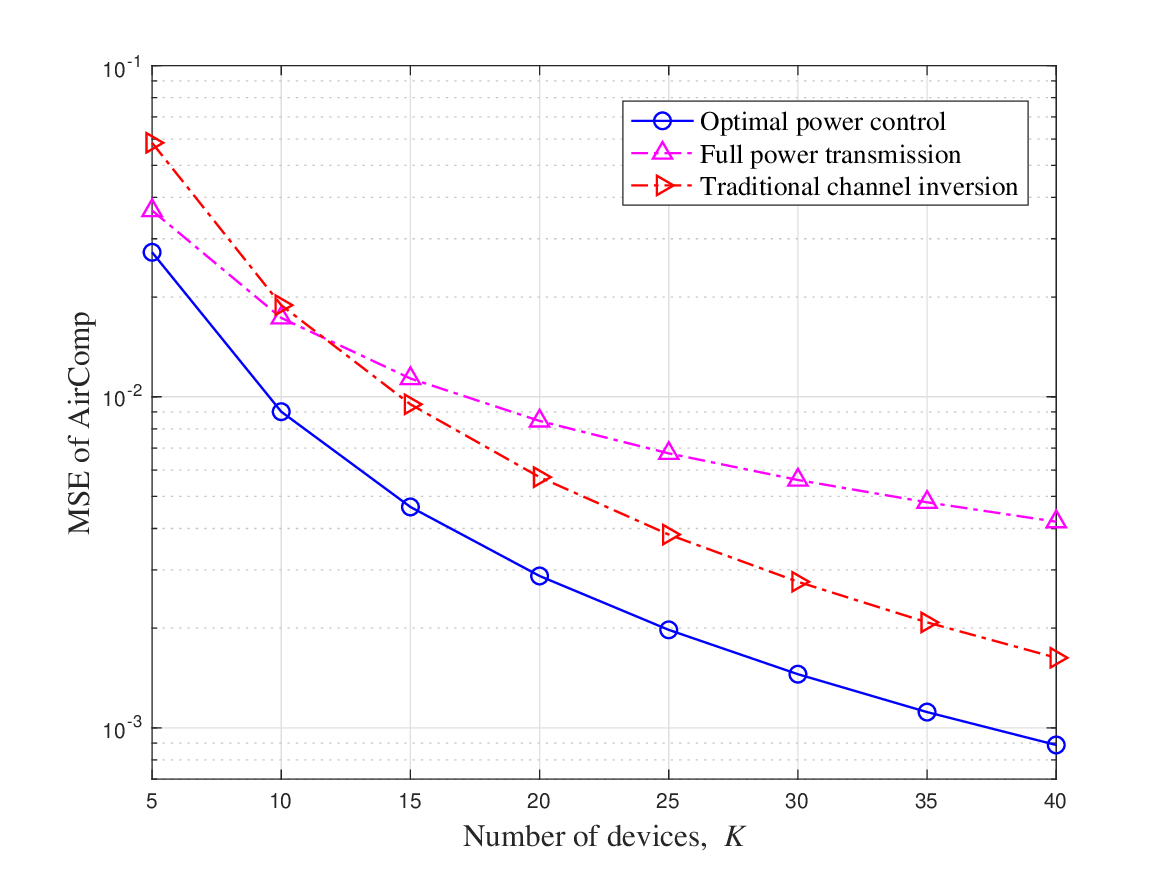}}
  \subfigure[Heterogenous expected receive SNR case.]
  {\label{fig:MSE_S_V_K_random}
\includegraphics[width=8cm]{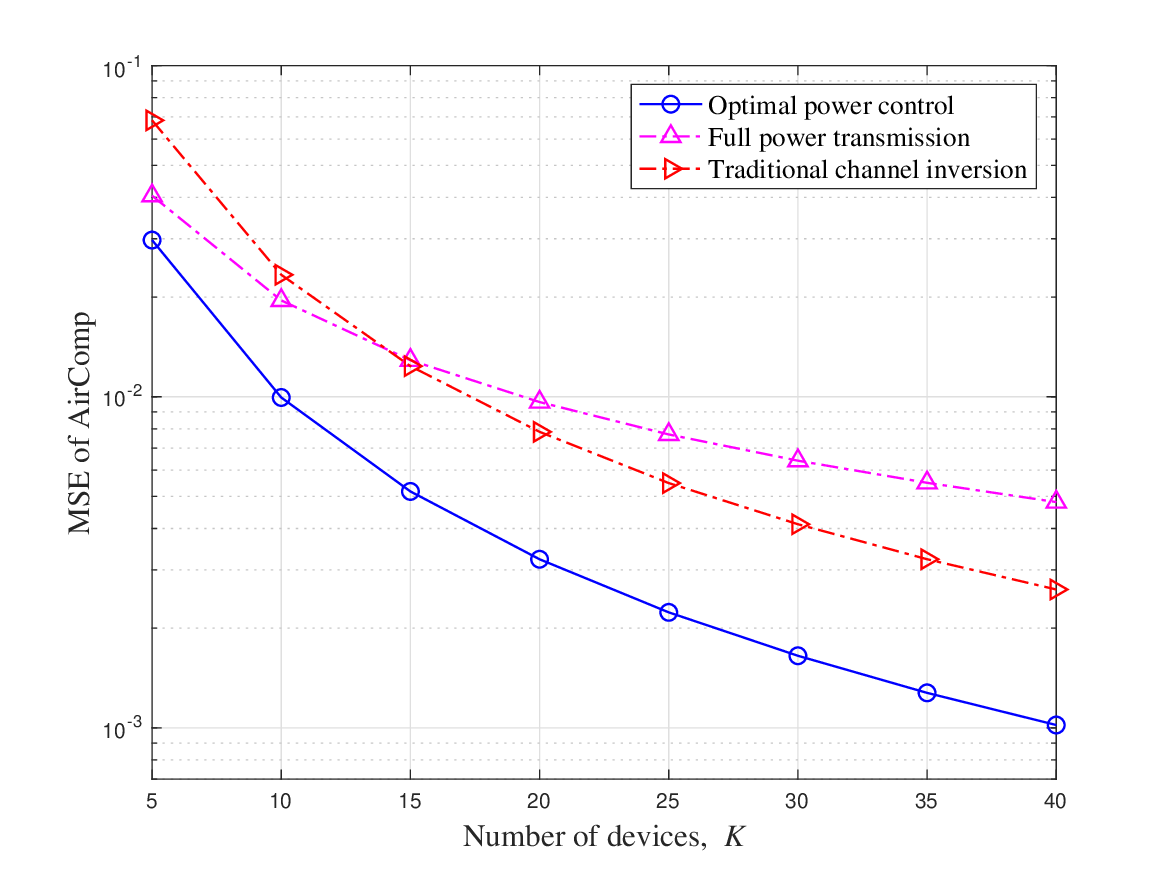}}
  \caption{Comparison of the number of devices on the MSE of AirComp in static channels, where the average receive SNR of the $K$ devices is set as $0$ dB.}
  \label{fig:MSE_S_V_K}
\vspace{-0.2cm}
\end{figure}

The MSE performance of AirComp in the static-channel case with independent observations is compared with that of two benchmark schemes introduced in the previous subsections.
The varying number of devices $K$ is considered in Figs.~\ref{fig:MSE_S_V_K_uni} and \ref{fig:MSE_S_V_K_random}, under the equal and unequal receive SNR scenarios, respectively, where the average receive SNR of the $K$ devices is set as $0$ dB.\footnote{ To ensure fair comparison, the total power budgets for the equal and unequal receive SNR scenarios are set to be the same. More specifically, in the equal-receive-SNR scenario, we have $\rho_k = 5$ dB, $\forall k\in K$; in unequal-receive-SNR scenario, the receive SNRs are set as $\rho_{5i-4} = 2.7$ dB, $\rho_{5i-3} = 4.5$ dB, $\rho_{5i-2} = 0$ dB, $\rho_{5i -1} = 5.4$ dB, $\rho_{5i} = 6.4$ dB, $\forall i = \{1,...,K/5\}$ with each value of $K$.}
Firstly, it is observed that the MSE achieved by all the three schemes decreases as $K$ increases, due to the fact that the FC receiver can aggregate more data for averaging.
Secondly, the proposed power control considerably outperforms both of the full power transmission and traditional-channel-inversion schemes throughout the whole regime of $K$.
The full power transmission outperforms the traditional-channel-inversion scheme when $K$ is small (i.e. $K\leq 12$ in Fig.~\ref{fig:MSE_S_V_K_uni}), but the performance compromises as $K$ increases, due to the lack of power adaptation to reduce the misalignment error.
Besides, it is interesting to observe that the MSE achieved under the equal-receive-SNR scenario in Fig.~\ref{fig:MSE_S_V_K_uni} is lower than that under the unequal-receive-SNR scenario in Fig.~\ref{fig:MSE_S_V_K_random}, although the total transmit power budgets of the $K$ devices are identical. This is because that the computation performance of AirComp is constrained by the device with the lowest power budget.

\begin{figure}
\centering
 \setlength{\abovecaptionskip}{-2mm}
\setlength{\belowcaptionskip}{-2mm}
    \includegraphics[width=8cm]{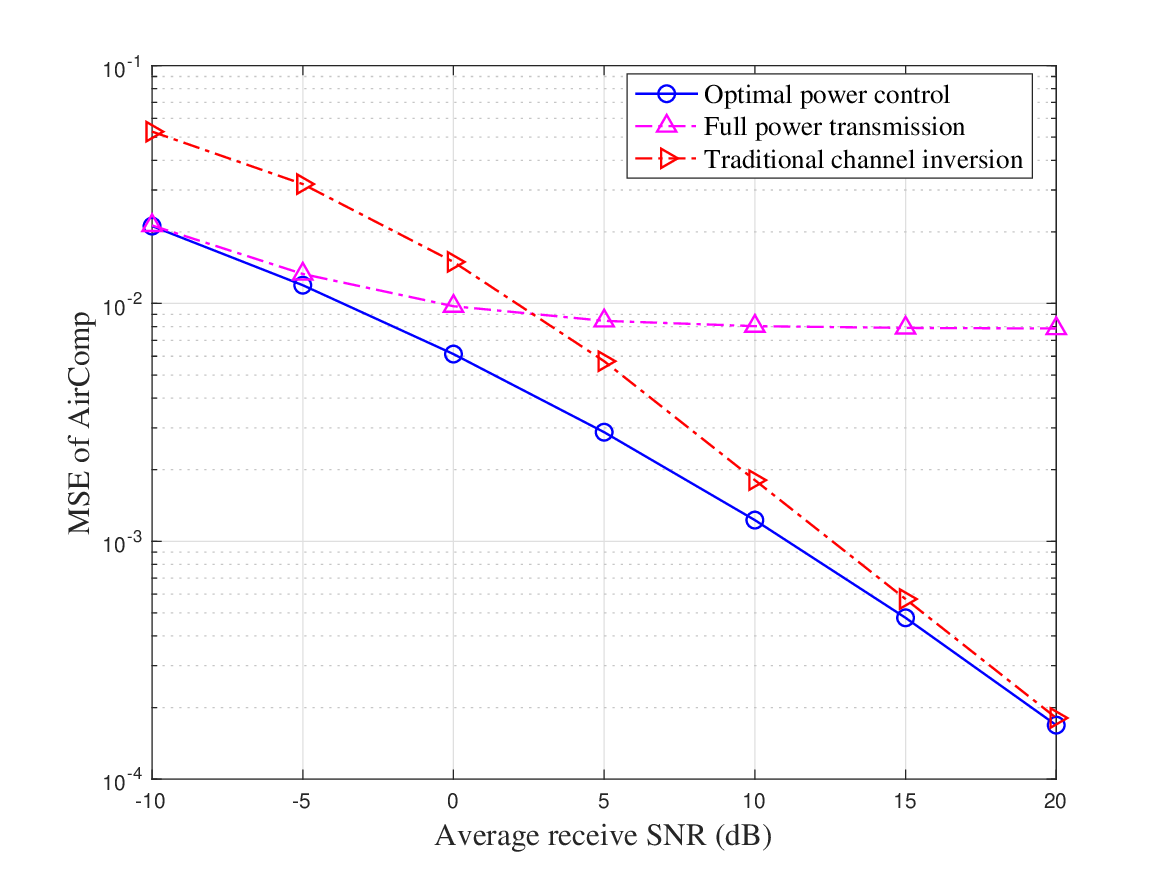}
\caption{Comparison of the average receive SNR on the MSE of AirComp in static channels with $K=20$.} \label{fig:MSE_S_v_P}
\vspace{-0.1cm}
\end{figure}
We further consider the varying average receive SNR in Fig.~\ref{fig:MSE_S_v_P}, which is equal for all devices with $K=20$.
It is observed that the achieved MSE by all schemes decreases with the average receive SNR increasing, and the proposed power control outperforms the other two benchmark schemes throughout the whole receive SNR regime.
With low receive SNR, the full power transmission has the same performance as the proposed power control, and both of them outperform the traditional-channel-inversion scheme. This is because that the full power transmission significantly suppresses the noise-induced error that is dominant for the MSE in this case.
As the receive SNR increases, the performance gap between the proposed power control and the full power transmission becomes large, while the traditional-channel-inversion scheme becomes close-to-optimal due to the perfect signal-magnitude alignment.
This coincides with our asymptotic analysis for static channels in Section \ref{static_asym}.

\vspace{-0.1cm}
\subsection{Performance of AirComp in Time-Varying Channels}

\begin{figure}[htbp]
  \centering
  \subfigure[Uniform expected receive SNR case.]
  {\label{fig:MSE_V_K_uni}\includegraphics[width=8cm]{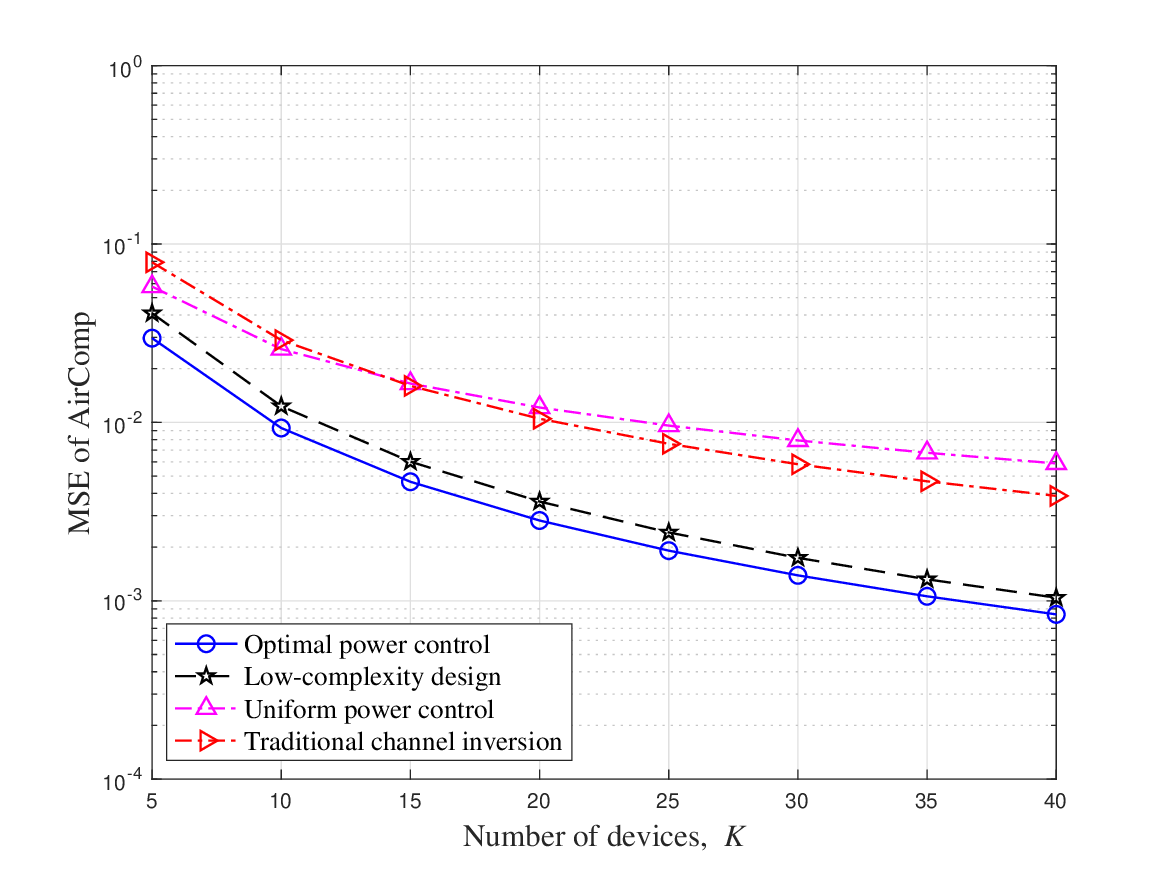}}
  \subfigure[Heterogenous expected receive SNR case.]
  {\label{fig:MSE_V_K_random}
\includegraphics[width=8cm]{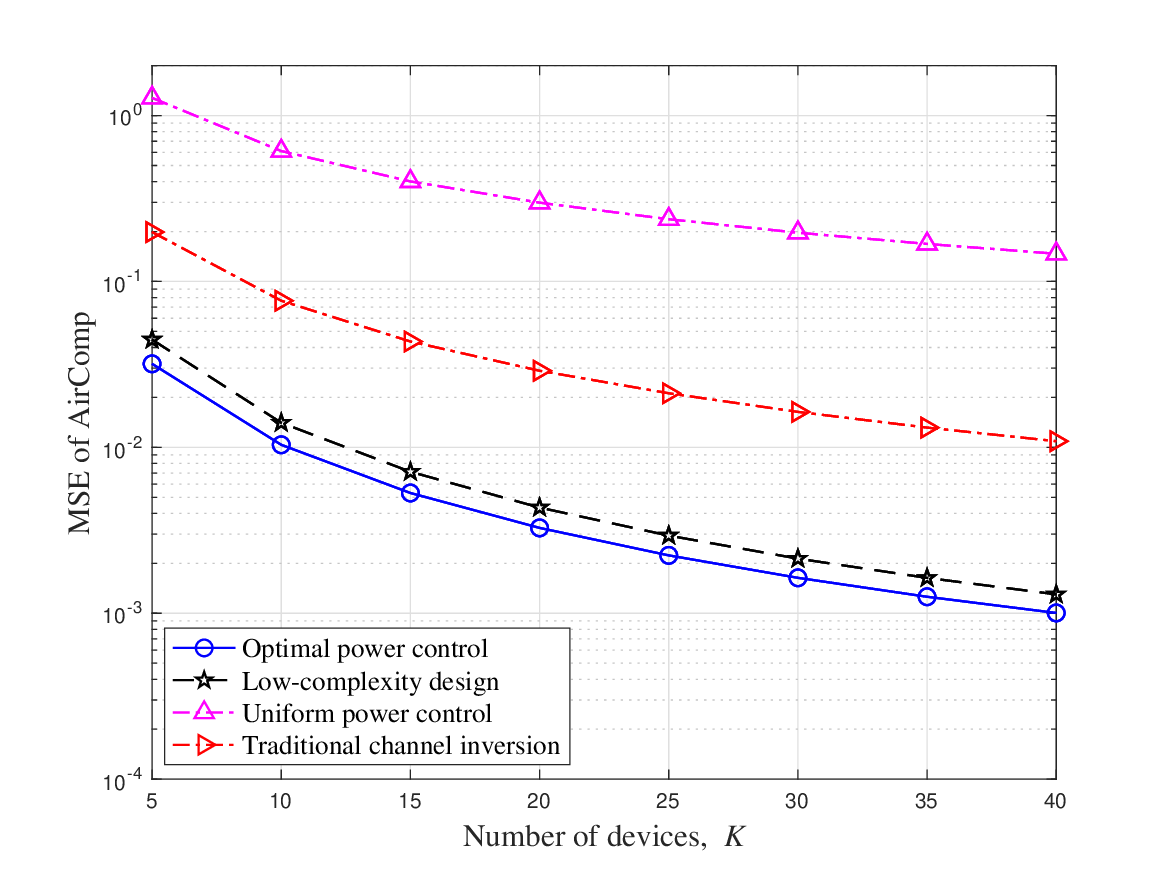}}
  \caption{Comparison of the number of devices on the MSE of AirComp in time-varying channels, where the average receive SNR is set to be $5$ dB.}
  \label{Fig:MSE_V_K}
\vspace{-0.1cm}
\end{figure}

For the case with time-varying channels, we compare the MSE performance of AirComp with independent observations with that of two benchmark schemes introduced before.
The varying number of devices $K$ is considered in Figs.~\ref{fig:MSE_V_K_uni} and \ref{fig:MSE_V_K_random} under the equal and unequal receive SNR scenarios, the same as those defined in Fig.~\ref{fig:MSE_S_V_K}.
Firstly, the MSE achieved by all schemes decreases as the number of devices $K$ increases.
Secondly, throughout the whole $K$ regime, the proposed power control outperforms the other three schemes while the low-complexity design is observed to be close-to-optimal, showing its effectiveness.
Besides, the uniform power control achieves better performance than the traditional-channel-inversion scheme, but the performance compromises quickly as $K$ increases due to the lack of power adaptation to the channel fluctuation.
Comparing Figs. \ref{fig:MSE_V_K_uni} and \ref{fig:MSE_V_K_random}, it is further observed that the MSE in the unequal-receive-SNR scenario is lower than that in the equal-receive-SNR scenario.
These observations are generally consistent with Fig.~\ref{fig:MSE_S_V_K} in the static channel case.
\begin{figure}
\centering
 \setlength{\abovecaptionskip}{-2mm}
\setlength{\belowcaptionskip}{-2mm}
    \includegraphics[width=8cm]{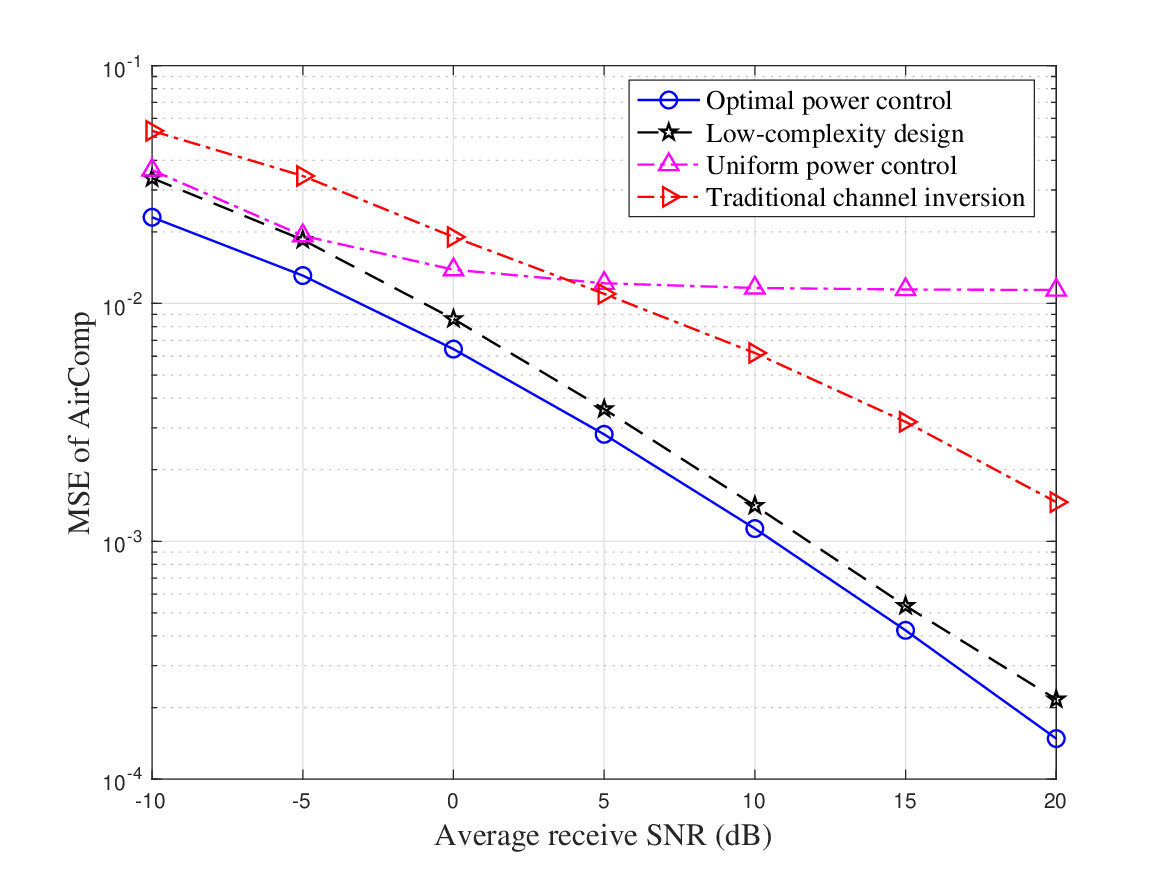}
\caption{Comparison of the average receive SNR on the MSE of AirComp in time-varying channels with $K=20$.} \label{fig:MSE_V_P}
\vspace{-0.1cm}
\end{figure}

Besides, the varying average receive SNR is considered in Fig.~\ref{fig:MSE_V_P}, which is uniform for all devices with $K= 20$.
It is observed that, for all schemes, the MSE decreases as the receive SNR increases.
The proposed power control performs consistently well throughout the whole receive SNR regime.
It is further observed that, the performance gap between the proposed and uniform power control is small in the low-receive-SNR regime while the gap becomes large at the high-receive-SNR regime. This shows the effectiveness of power control optimization, especially in the high-receive-SNR regime.
Particularly, when the receive SNR is $20$ dB, both the proposed power control and low-complexity design achieve up to two orders of magnitude AirComp error reduction, as compared to the uniform power control.
It is interesting to emphasize that such phenomena is in sharp contrast to the power control for rate maximization in conventional wireless communication systems over fading channels (e.g., single-user point-to-point channels), where the adaptive power control (e.g., water-filling \cite{Goldsmith1997}) is more crucial in the low to moderate SNR regimes.

\vspace{-0.3cm}
\subsection{Effect of Imperfect CSI}

\begin{figure}
\centering
 \setlength{\abovecaptionskip}{-2mm}
\setlength{\belowcaptionskip}{-2mm}
    \includegraphics[width=8cm]{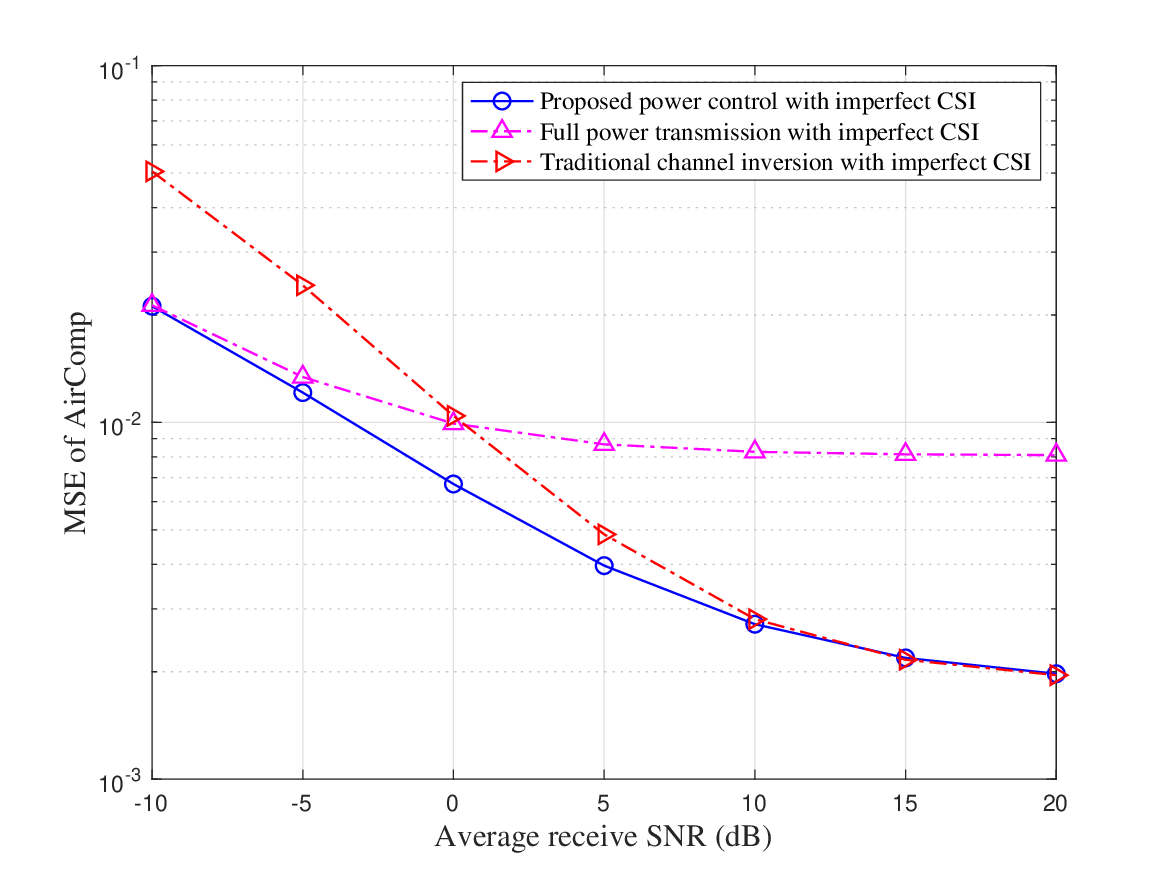}
\caption{MSE performance of AirComp with the imperfect CSI.} \label{fig:MSE_S_v_P_ImperfectCSI}
\vspace{-0.1cm}
\end{figure}


In practice, due to the channel estimation error, the CSI can be imperfect at both the device transmitters and the FC receiver. To gain more insights, we show the the MSE performance of AirComp with the imperfect CSI by considering the static-channel case with $K = 20$, shown in Fig. \ref{fig:MSE_S_v_P_ImperfectCSI}, for which the system parameters are set same as for Fig. \ref{fig:MSE_S_v_P}. In the simulation, we model the imperfect CSI by considering $h_k = \bar h_k + \epsilon_k, \forall k\in \cal K$, where $\bar h_k$ denotes the estimated CSI and $\epsilon_k$ denotes the estimation error at device $k\in \cal K$, which is around $50\%$ of the ground truth channel gain.
By comparing Fig. \ref{fig:MSE_S_v_P_ImperfectCSI} versus Fig. \ref{fig:MSE_S_v_P}, it is observed that in the low-receive-SNR regime (e.g., -5 dB), the MSE values with imperfect CSI in Fig. \ref{fig:MSE_S_v_P_ImperfectCSI} are similar as those with perfect CSI in Fig. \ref{fig:MSE_S_v_P}, especially for the proposed power control and full power transmission. This is due to the fact that in this case, the MSE performance is dominated by the noise and thus the channel estimation error has negligible influence. By contrast, in the high-receive-SNR regime (e.g., 20 dB), the MSE values with imperfect CSI in Fig. \ref{fig:MSE_S_v_P_ImperfectCSI} are much higher than those with perfect CSI in Fig. \ref{fig:MSE_S_v_P}. This is because that the MSE performance is dominated by the signal-misalignment error, such that the channel estimation error becomes significant.
The observation aligns with the intuition that the MSE performance is more sensitive to the accuracy of channel estimation in the high-receive-SNR regime.

\vspace{-0.1cm}
\subsection{Effect of Correlated Observations}

\begin{figure}
\centering
 \setlength{\abovecaptionskip}{-2mm}
\setlength{\belowcaptionskip}{-2mm}
    \includegraphics[width=8cm]{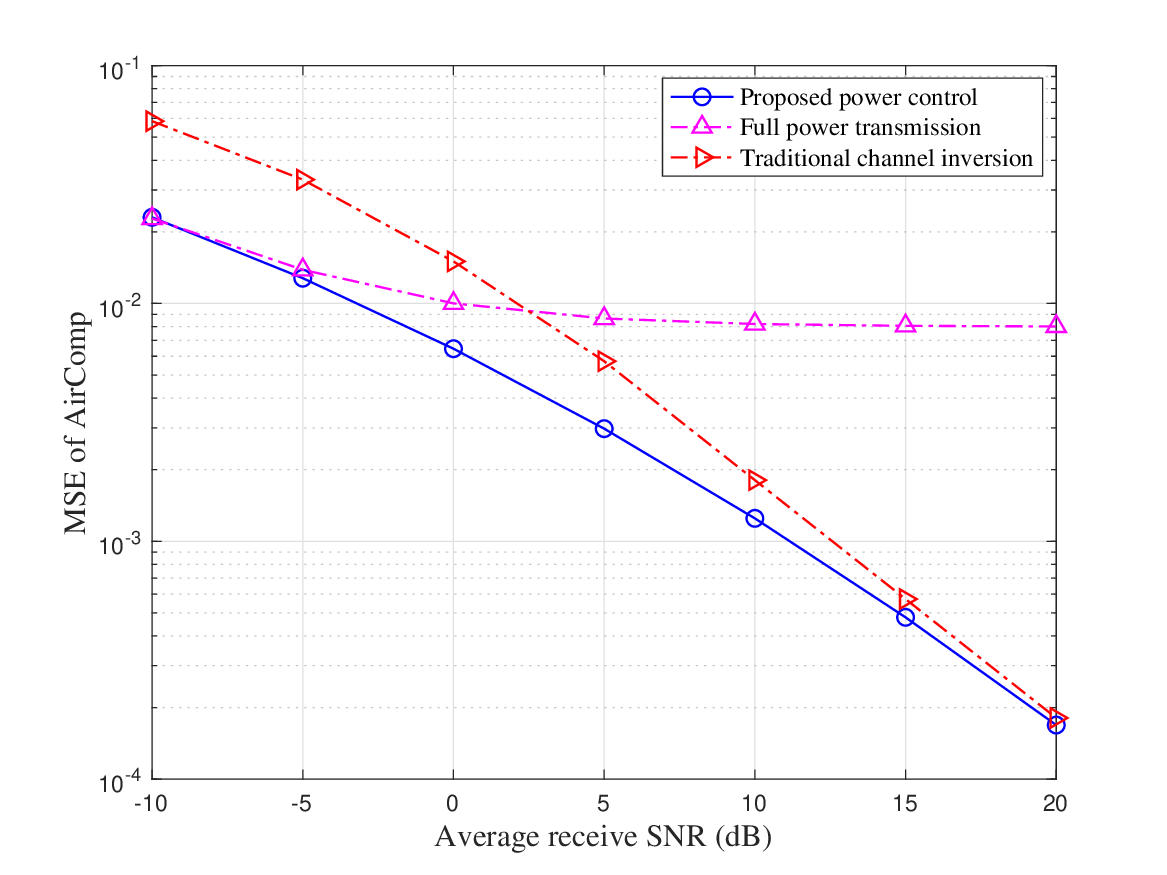}
\caption{Comparison of the average receive SNR on the MSE of AirComp in static channels with $K=20$, where the observations are correlated.} \label{fig:MSE_S_v_P_cor}
\vspace{-0.1cm}
\end{figure}

\begin{figure}
\centering
 \setlength{\abovecaptionskip}{-2mm}
\setlength{\belowcaptionskip}{-2mm}
    \includegraphics[width=8cm]{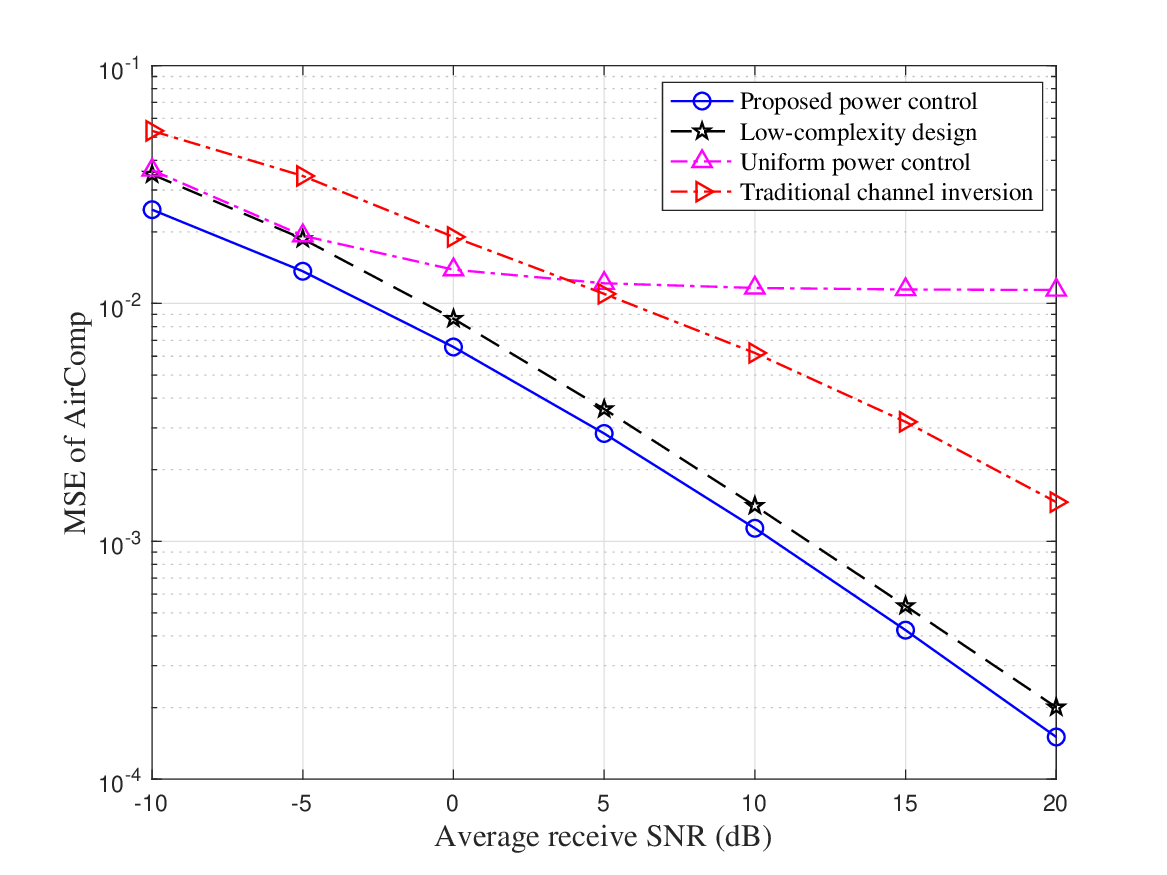}
\caption{Comparison of the average receive SNR on the MSE of AirComp in time-varying channels with $K=20$, where the observations are correlated.} \label{fig:MSE_V_P_cor}
\vspace{-0.1cm}
\end{figure}

Finally, we evaluate the the exact MSE performance (i.e. $\widehat {\rm MSE}(t)$) of AirComp with correlated observations targeting static-channel and time-varying-channel cases respectively in Figs. \ref{fig:MSE_S_v_P_cor} and \ref{fig:MSE_V_P_cor}, for which the system parameters are set as the same as those for Figs. \ref{fig:MSE_S_v_P} and \ref{fig:MSE_V_P}, respectively.
 In this case, the sensing values at devices are assumed to follow the multivariate Gaussian distribution with mean $\bm \Theta_{ \textbf{z}}$ and covariance matrix $\bm \Sigma_{ \textbf{z}}$, and thus the PDF is given by
 \begin{align*}
 	p_z=\frac{1}{(2\pi)^{\frac{K}{2}}|\bm \Sigma_{ \textbf{z}} |^{\frac{1}{2}} }\exp\left(-\frac{1}{2}(\textbf{z}-\bm \Theta_{ \textbf{z}})^{\rm T}\bm \Sigma_{ \textbf{z}}^{-1}(\textbf{z}-\bm \Theta_{ \textbf{z}})\right).
 \end{align*}
Note that the covariance matrix can be decomposed into $\bm \Sigma_{ \textbf{z}}=\textbf{D}^{\frac{1}{2}} \textbf{C}_\textbf{z} \textbf{D}^{\frac{1}{2}}$, where $\textbf{D}$ is a diagonal matrix with the diagonal elements collecting the variances of the elements of $\textbf{z}$, which are set to be 1, and $\textbf{C}_\textbf{z}$ is a correlation matrix that is assumed to follow the linear exponent autoregressive structure \cite{Covarianve2010}:
 \begin{align}
 	\textbf{C}_\textbf{z}=\left[
 	\begin{matrix}
     1 & c & c^2& \cdots &c^{k-1}\\
      c & 1& c& \cdots &c^{k-2}\\
       \cdots & \cdots & \cdots & \quad & \cdots\\
        c^{k-1} & c^{k-2} & c^{k-2} & \cdots &1
\end{matrix}\right],
 \end{align}
 where $c\in [0,1)$ represents the level of correlation, and we set $c=0.1$ here.
We substitute the optimized power control policy derived in Theorems \ref{eta_K_static} and \ref{P1_opt_sol} into \eqref{MSE_t1} to show the resultant MSE performance in static-channel and time-varying-channel cases, respectively.
By comparing Fig. \ref{fig:MSE_S_v_P_cor} with Fig. \ref{fig:MSE_S_v_P}, it is observed that the trend of the exact MSE performance for static-channel case in Fig. \ref{fig:MSE_S_v_P_cor} follows that in Fig. \ref{fig:MSE_S_v_P}. Similar observations can be also made for the time-varying-channel case by comparison between Figs. \ref{fig:MSE_V_P_cor} and \ref{fig:MSE_V_P}.

\vspace{-0.1cm}
\section{Concluding Remarks}\label{sec_con}\vspace{-0.1cm}
This work solved the power control optimization problem for an AirComp system operating over a fading MAC.
We presented the power-control policy for different cases.
In the case with static channels, the derived power control exhibited a threshold-based structure, where a device should transmit either using channel-inversion policy or with full power depending on a derived quality measure of the device.
In the case with time-varying channels, the proposed power-control policy had a regularized channel inversion structure, where the regularization balances the tradeoff between the signal-misalignment error reduction and noise suppression.
Furthermore, for the special case that there exists only one power-limited device in AirComp system, this device should adopt channel-inversion water-filling power control over fading states, while other devices with unlimited power budgets should transmit with channel-inversion power control.
Remarkable performance gain in terms of AirComp accuracy was observed in the comparison with heuristic designs, necessitating the optimal power control for reliable AirComp.

This current work also opens several directions for further investigation, as briefly discussed in the following to motivate future work.
\begin{itemize}
\item One direction is to design practical online power-control policy for energy harvesting AirComp, in which the sensors are each powered by energy harvesting devices such as solar and wind energy sources (instead of fixed power supply in this paper). The key challenge lies in the power adaption under the energy harvesting constraints that are imposed due to the randomness and intermittence of energy arrivals \cite{EH2015}.
 \item Another interesting direction is to explore AirComp cooperation or interference control for large-scale AirComp over multiple FCs in IoT networks, which needs to balance the radio resource allocation at FCs for interference suppression and computation error reduction.
\end{itemize}

\vspace{-0.1cm}
\appendix
\vspace{-0.2cm}
\subsection{ Proof of Lemma~\ref{opt_noactive_lemma}}\label{opt_noactive_proof}
Consider the case with $\eta\in{\cal F}_0$, for which the problem \eqref{static_K_eta} (or \eqref{F_k}) can be expressed as
 \begin{align}\label{lemma_1_F0}
\min_{\eta\in {\cal F}_0} F_0(\eta) = \min_{0\le \eta \le \bar P_1 |h_1|^2}
 ~ \frac{\sigma^2}{\eta}.
\end{align}
It is evident that the optimal solution to problem \eqref{lemma_1_F0} is $\eta _0^* = \bar P_1|h_1|^2$,
therefore, $\eta _0^* =  \bar P_1|h_1|^2$ can achieve a lower MSE value than any $\eta <  \bar P_1|h_1|^2$. As a result, it must hold that $\eta^*\ge\bar{P}_1|h_1|^2$ for problem \eqref{static_K_eta}. Accordingly, it follows from \eqref{K_device_N1_Pk_opyt} that device $1$ should always transmit with full power, i.e., $p_1^*=\bar P_k$. This completes the proof.
\vspace{-0.1cm}
\subsection{ Proof of Lemma~\ref{de_in_function_lemma}}\label{de_in_function_proof}

With any given $k\in \cal K$, the first-order derivative of the objective function $F_k(\eta)$ w.r.t. $\eta$ is derived as
 \begin{align}\label{F_k_deri}
F_k^{'}(\eta)=\frac{ \sqrt{\eta} \sum_{i=1}^k\bar{P}_i|h_{i}|^2- \sum_{i=1}^k \bar P_i|h_{i}|^2-\sigma^2 } {\eta^2 }.
\end{align}
It is easy to check that the equation $F_k^{'}(\eta) = 0$ has a unique solution given by
 \begin{align}
\tilde \eta_{k}=\left( \frac{\sigma^2+ \sum_{i=1}^{k} \bar{P}_{i}|h_{i}|^2}{\sum_{i=1}^{k}\sqrt{\bar{P}_{i}}|h_{i}|}\right)^2.
\end{align}
It also follows that when $\eta\geq \tilde \eta_{k}$, we have $F_k^{'}(\eta)\geq0$; while when $\eta\leq \tilde \eta_{k}$, we have $F_k^{'}(\eta)\leq0$. Therefore, the function $F_k(\eta)$ is  monotonically decreasing over $\eta \in [0,\tilde \eta_k]$ and monotonically increasing over $\eta \in [\tilde \eta_k, \infty)$.
In this case, by comparing the achieved function values of $F_k(\eta)$ at the boundary points of $\bar P_k|h_{k}|^2$ and $\bar P_{k+1}|h_{k+1}|^2$, and that at the stationary point $\tilde \eta_{k}$, we have the optimal solution $\eta_k^*$ to problem \eqref{F_k} shown in \eqref{static_eta_k}.
Therefore, Lemma~\ref{de_in_function_lemma} is proved.

\vspace{-0.1cm}
\subsection{ Proof of Theorem~\ref{eta_K_static}}\label{eta_K_static_proof}
\vspace{-0.1cm}
Note that the threshold-based optimal power-control policy in \eqref{static_p_lemma} can be easily obtained by combining \eqref{K_device_N1_Pk_opyt} and $\eta^*$.
Therefore, to complete the proof, it only remains to show $\eta^*=\eta^*_{k^*}=\tilde \eta_{k^*}$. Based on \eqref{static_eta_k}, we can prove this by equivalently showing that $\eta_{k^*}^* \neq \bar{P}_{k^*}|h_{k^*}|^2$ and $\eta_{k^*}^* \neq \bar{P}_{k^*+1}|h_{k^*+1}|^2$. Towards this end, we first have the following lemma:
\begin{lemma}\label{eta_P_verse_lemma}\emph{
For any $k\in{\cal K}$, we have $\bar{P}_{k}|h_{k}|^2\leq\tilde \eta_{k}  \iff  \bar{P}_{k}|h_{k}|^2\leq\tilde \eta_{k-1}$ and $\bar{P}_{k}|h_{k}|^2\geq\tilde \eta_{k}  \iff  \bar{P}_{k}|h_{k}|^2\geq\tilde \eta_{k-1}$, where we define $\tilde \eta_0\to \infty$. 
 }
\end{lemma}
\vspace{-0.3cm}
\begin{IEEEproof}
	See Appendix~\ref{eta_P_verse_proof}.
\end{IEEEproof}
First, we show $\eta_{k^*}^* \neq \bar{P}_{k^*}|h_{k^*}|^2$ by contradiction. Suppose that $\eta^* = \eta_{k^*}^*=\bar{P}_{k^*}|h_{k^*}|^2$, which is the globally optimal solution to problem \eqref{static_K_eta}.
As $\bar P_{k^*} |h_{k^*}|^2$ lies in both regions of ${\cal F}_{k^*-1}$ and ${\cal F}_{k^*}$, it is evident that  $\bar P_{k^*} |h_{k^*}|^2$ is also optimal for both $\min_{\eta\in {\cal F}_{k^*-1}} F_{k^*-1}(\eta)$ and $\min_{\eta\in {\cal F}_{k^*}} F_{k^*}(\eta)$ in \eqref{F_k} for $k = k^*-1$ and $k^*$, respectively.
Therefore, based on \eqref{static_eta_k}, it must follow that $\bar P_{k^*} |h_{k^*}|^2 \le \tilde \eta _{k^*-1}$ and  $\bar P_{k^*} |h_{k^*}|^2 \ge \tilde \eta_{k^*}$, respectively.
Notice that $\tilde \eta _{k^*-1} \neq \tilde \eta_{k^*}\neq \bar P_{k^*} |h_{k^*}|^2$, and therefore, the two equations contradict with Lemma \ref{eta_P_verse_lemma}. As a result, the presumption of $\eta_{k^*}^* = \bar P_{k^*} |h_{k^*}|^2$ cannot be true. Therefore, we have $\eta_{k^*}^* \neq \bar P_{k^*} |h_{k^*}|^2$.

Next, we can also show $\eta_{k^*}^* \neq \bar{P}_{k^*+1}|h_{k^*+1}|^2$ via a similar proof technique based on contradiction, for which the details are omitted for brevity.

Therefore, combining the above two cases together with \eqref{static_eta_k}, it must hold that $\eta^*=\eta_{k^*}^*=\tilde \eta_{k^*}$. Therefore, Theorem \ref{eta_K_static} is proved.

\vspace{-0.1cm}
\subsection{ Proof of Lemma~\ref{eta_P_verse_lemma}}\label{eta_P_verse_proof}
\vspace{-0.1cm}
Essentially, Lemma~\ref{eta_P_verse_lemma} says that $\bar{P}_{k}|h_{k}|^2-\tilde \eta_{k}$ and $\bar{P}_{k}|h_{k}|^2-\tilde \eta_{k-1}$ have the same sign.
First, we consider $\bar{P}_{k}|h_{k}|^2-\tilde \eta_{k}$. It follows that
\begin{align*}
&\bar{P}_{k}|h_{k}|^2\!-\!\tilde \eta_{k}\!\\
&\! \! =A_k\left[\!\frac{\sum_{i=1}^{k}\! \!\sqrt{\bar{P}_{i}}|h_{i}|\! (\! \sqrt{\bar{P}_{k}}|h_{k}|\!-\!\sqrt{\bar{P}_{i}}|h_{i}|\! )\!-\!\sigma^2}{\sum_{i=1}^{k}\! \! \sqrt{\bar{P}_{i}}|h_{i}|}\!\right].\!
\end{align*}
Note that $A_k\triangleq \!\sqrt{\bar{P}_{k}}|h_{k}|\!\! +\! \!\left( \frac{\sigma^2\!+\! \sum_{i=1}^{k} \! \bar{P}_{i}|h_{i}|^2\! }{\! \sum_{i=1}^{k}\! \! \sqrt{\bar{P}_{i}}|h_{i}|}\! \right)$ and  $\sum_{i=1}^{k}\!\sqrt{\bar{P}_{i}}|h_{i}|(\sqrt{\bar{P}_{k}}|h_{k}|\!-\!\sqrt{\bar{P}_{i}}|h_{i}|)\!=\!\sum_{i=1}^{k-1}\!\sqrt{\bar{P}_{i}}|h_{i}|\!\left(\!\!\sqrt{\bar{P}_{k}}|h_{k}|\!-\!\sqrt{\bar{P}_{i}}|h_{i}|\!\right)\!$. By defining
\begin{align}\label{J_k}
J(k)\triangleq \sum_{i=1}^{k-1}\sqrt{\bar{P}_{i}}|h_{i}|\left(\sqrt{\bar{P}_{k}}|h_{k}|-\sqrt{\bar{P}_{i}}|h_{i}|\right)-\sigma^2,
\end{align}
with $J(1)\triangleq -\sigma^2$, we have
\begin{align}
		&\bar{P}_{k}|h_{k}|^2-\tilde \eta_{k}\notag\\
		&\! =\! \left[\! \!  \sqrt{\bar{P}_{k}}|h_{k}|\! \! +\! \! \left(\!  \frac{\sigma^2\! +\!  \sum_{i=1}^{k}\!  \bar{P}_{i}|h_{i}|^2}{\sum_{i=1}^{k}\!\! \sqrt{\bar{P}_{i}}|h_{i}|}\! \right)\! \right]\!\!  \left(\! \!   \frac{J(k)}{\sum_{i=1}^{k}\sqrt{\bar{P}_{i}}|h_{i}|}\! \right).\label{F_opt_k}
\end{align}
It thus follows from \eqref{F_opt_k} that the sign of $\bar{P}_{k}|h_{k}|^2-\tilde \eta_{k}$ is same as that of $J(k)$. We then have
\begin{align}
	&\bar{P}_{k}|h_{k}|^2\leq\tilde \eta_{k} \iff J(k)\leq 0,~ \forall k\in{\cal K},\label{J1}\\
	&\bar{P}_{k}|h_{k}|^2\geq\tilde \eta_{k} \iff J(k)\geq 0,~ \forall k\in{\cal K}.\label{J2}
\end{align}
Next, we consider $\bar{P}_{k}|h_{k}|^2-\tilde \eta_{k-1}$. Similarly as for \eqref{F_opt_k}, it follows based some simple manipulations that
\begin{align}
		&\bar{P}_{k}|h_{k}|^2-\tilde \eta_{k-1}\notag\\
		&\!\!=\!\!\left[\! \!\sqrt{\bar{P}_{k}}|h_{k}|\!+\!\!\left(\! \frac{\sigma^2\!+ \!\sum_{i=1}^{k-1} \!\!\bar{P}_{i}|h_{i}|^2\!}{\!\sum_{i=1}^{k-1}\!\sqrt{\bar{P}_{i}}|h_{i}|}\!\right)\!\right]\!\left(\! \frac{J(k)}{\sum_{i=1}^{k-1}\!\sqrt{\bar{P}_{i}}|h_{i}|}\!\right),\!\label{F_opt_kminus1}
\end{align}
 where we define $\tilde \eta_0\to \infty$.
Thus, the sign of $\bar{P}_{k}|h_{k}|^2-\tilde \eta_{k-1}$ is also same as that of $J(k)$, i.e., \begin{align}
&\bar{P}_{k}|h_{k}|^2\leq\tilde \eta_{k-1} \iff J(k)\leq 0,~ \forall k\in{\cal K},\label{J3}\\
	&\bar{P}_{k}|h_{k}|^2\geq\tilde \eta_{k-1} \iff J(k)\geq 0,~ \forall k\in{\cal K}.\label{J4}
\end{align}
By combining \eqref{J1}, \eqref{J2}, \eqref{J3}, and \eqref{J4},
 the proof of Lemma~\ref{eta_P_verse_lemma} is thus completed.


\vspace{-0.1cm}
\subsection{ Proof of Lemma~\ref{opt_eta_static_ordering_lemma}}\label{opt_eta_static_ordering_proof}
\vspace{-0.1cm}
Notice that property 1) follows directly based on \eqref{opt_eta_static}
 in Theorem 1. Therefore, we only need to show the remaining two properties in the following.

First, consider property 2). As in property 1), we have $\bar{P}_{k^*}|h_{k^*}|^2\leq\tilde \eta_{k^*-1}$. It thus follows from \eqref{J3} that $J(k^*)\leq0$.
Notice that based on \eqref{J_k}, we have
\begin{align}
	J(1) \le \cdots\le J(k)\le\cdots\le J(K).  \label{B}
\end{align}
Therefore, based on $J(k^*) \le 0$, we have $J(k) \le 0, \forall k\in\{1,...,k^*\}$.
Therefore, by using \eqref{J3} again, we have $\bar{P}_{k}|h_{k}|^2- \tilde \eta_{k-1}\leq 0$, $\forall k\in\{1,\cdots, k^*\}$, with $\tilde \eta_{(0)}\to \infty$.

Similarly, based on $\bar{P}_{k^*+1}|h_{k^*+1}|^2\geq\tilde \eta_{k^*}$, we can show that $\bar{P}_{k}|h_{k}|^2-\tilde \eta_{k-1}\geq 0$, $\forall k\in\{k^*+1,\cdots, K\}$.
Hence, property 2) is proved.

Next, consider property 3). Towards this end, in the following, we show $\tilde \eta_{k}\leq\tilde \eta_{k-1}, \forall k\in\{1,\cdots,k^*\}$ and $\tilde \eta_{k}\geq\tilde \eta_{k-1}, \forall k\in\{k^*+1,\cdots,K\}$.
It is easy to show that
\begin{align}
		&\tilde \eta_{k}- \tilde\eta_{k-1}\notag\\
		&\!=\!\!\left(\!\sqrt{\tilde \eta_{k}}\!\!+\!\sqrt{\tilde\eta_{k-1}}\!\right)\!\!\left[\!  \frac{J(k)\sqrt{\bar{P}_{k}}|h_{k}| \!}{\!\left(\sum_{i=1}^{k}\!\sqrt{\bar{P}_{i}}|h_{i}|\!\right)\!\left(\sum_{i=1}^{k-1}\!\sqrt{\bar{P}_{i}}|h_{i}|\!\right)\!}\!\right].\label{eta_k_property3}
\end{align}
Recall that $\bar{P}_{k^*}|h_{k^*}|^2\leq\tilde \eta_{k^*}$ and $ \bar{P}_{k^*+1}|h_{k^*+1}|^2\geq\tilde \eta_{k^*}$ in property 1), and it thus follows from \eqref{J1} and \eqref{J4} that  $J(k^*)\leq 0$ and $J(k^*+1)\geq 0$, respectively.
It is easy to know that $\tilde \eta_{k}\leq\tilde \eta_{k-1}, \forall k\in\{1,\cdots,k^*\}$ and $\tilde \eta_{k}\geq\tilde \eta_{k-1}, \forall  k\in\{k^*+1,\cdots,K\}$.
Therefore, we have $k^* = \arg \min_{k\in\cal K} \tilde \eta_{k}$.

By combing the three properties, this lemma is finally proved.

\vspace{-0.5cm}
\subsection{Proof of Theorem~\ref{general_K_activenum_proof}}\label{general_K_activenum}
\vspace{-0.3cm}
We use contradiction to prove this Theorem. Suppose that the transmit power of any one particular device $i\in\cal K$ is not used up. In this case,  we have $\mu^*_i = 0$ based on the complementary slackness condition $\mu^*_i\left(\mathbb{E}_{\bm\nu}(p_i^*(\bm\nu))-\bar{P}_i\right)=0$. It thus follows from \eqref{opt_Pkv} that device $i$ transmits with channel-inversion power control, i.e.,
\begin{align}\label{P1_ray}
	p_i(\bm\nu)=\frac{\eta(\bm\nu)}{|h_{i}|^2},
\end{align}
where $\eta(\bm\nu) > 0$ holds in general. With Rayleigh fading, it happens that the wireless channel gains can be sufficiently small, thus leading to sufficiently large transmit power based on \eqref{P1_ray}. In this case we have
\begin{align}
	\mathbb{E}_{\bm\nu}[p_i^*(\bm\nu)] \to \infty.
\end{align}
This thus contradicts the average transmit power constraint at device $i$. As a result, the presumption $\mu_i = 0$ cannot be true. With $\mu_i > 0$, we have that all average power constraints should be tight in Rayleigh fading channel case, i.e., $\mu_k>0, \forall k\in\cal K$. This thus completes the proof of this theorem.

\vspace{-0.1cm}
\subsection{Proof of Lemma~\ref{Specific_K_lemma}}\label{Specific_K_proof}
\vspace{-0.1cm}
As problem \eqref{two_case_P1_OPT} is convex and satisfies the Slater's condition, we can solve this problem by applying the KKT conditions \cite{Boyd2004}.
The Lagrangian of problem \eqref{two_case_P1_OPT} is given by
\begin{align*}
{\cal L}_1\!=\!\mathbb{E}_{\bm\nu}\!\left[\frac{\sigma^2}{\sigma^2\!+p_{1}\!(\bm\nu)\!|h_{1}|^2}+\mu_1p_1(\bm\nu)\!\right]\!-\!\mu_1 \bar{P}_1\!-\!\mathbb{E}_{\bm\nu}[a_{\bm\nu}p_1(\bm\nu)],\!
\end{align*}
where $a_{\bm\nu}$ and $\mu_1$ denote the non-negative Lagrange multipliers associated with $p_1(\bm\nu)\geq 0, \forall \bm\nu $, and  \eqref{P3_power}, respectively.
Let $\{p_1^{\rm opt}(\bm\nu)\}$ denote the optimal solution to problem \eqref{two_case_P1_OPT}, as well as $\{a_{\bm\nu}^{\rm opt}\}$ and $\mu_1^{\rm opt}$ denote the optimal Lagrange multipliers.
Thus they must satisfy the KKT conditions as follows:
\begin{align}
&a_{\bm\nu} \ge 0, ~\mu_1 \ge 0,~ p_1(\bm\nu)\ge 0,~\forall \bm\nu\\
&a_{\bm\nu}^{\rm opt}p_1^{\rm opt}(\bm\nu)=0,~\forall \bm\nu\label{eq.kkt1} \\
&\mu_1^{\rm opt}(\mathbb{E}_{\bm\nu}(p_1^{\rm opt}(\bm\nu))- \bar{P}_1)=0\label{eq.kkt2}\\
&\mu_1^{\rm opt}-\frac{\sigma^2 |h_{1}|^2}{(\sigma^2+p_1^{\rm opt}(\bm\nu)|h_{1}|^2) ^2} -a_{\bm\nu}^{\rm opt}=0,~\forall \bm\nu, \label{eq.kkt3}
\end{align}
where \eqref{eq.kkt1} and \eqref{eq.kkt2} denote the complementary slackness conditions, and \eqref{eq.kkt3} is obtained from the first-order derivative conditions of ${\cal L}_1$ w.r.t. $p_1(\bm\nu)$'s.
Combining \eqref{eq.kkt1} and \eqref{eq.kkt3}, we obtain the optimal $p_1^{\rm opt}(\bm\nu)$ in  \eqref{K2_opt_p_1n}.
Notice that if $\mu_1^{\rm opt}=0$, $p_1^{\rm opt}(\bm\nu)$ approaches infinity and violates \eqref{P3_power}.
Hence, it must hold $\mu_1^{\rm opt}>0$ and \eqref{eq.kkt2} becomes
\begin{align}\label{P3_comp_P1}
	\mathbb{E}_{\bm\nu}[p_1^{\rm opt}(\bm\nu)]-\bar{P}_1=0,
\end{align}
based on which we can use a bisection search to obtain $\mu_1^{\rm opt}$.
Thus, Lemma~\ref{Specific_K_lemma} is proved.


\begin{IEEEbiography}[{\includegraphics[width=1in,height=1.25in,clip,keepaspectratio]{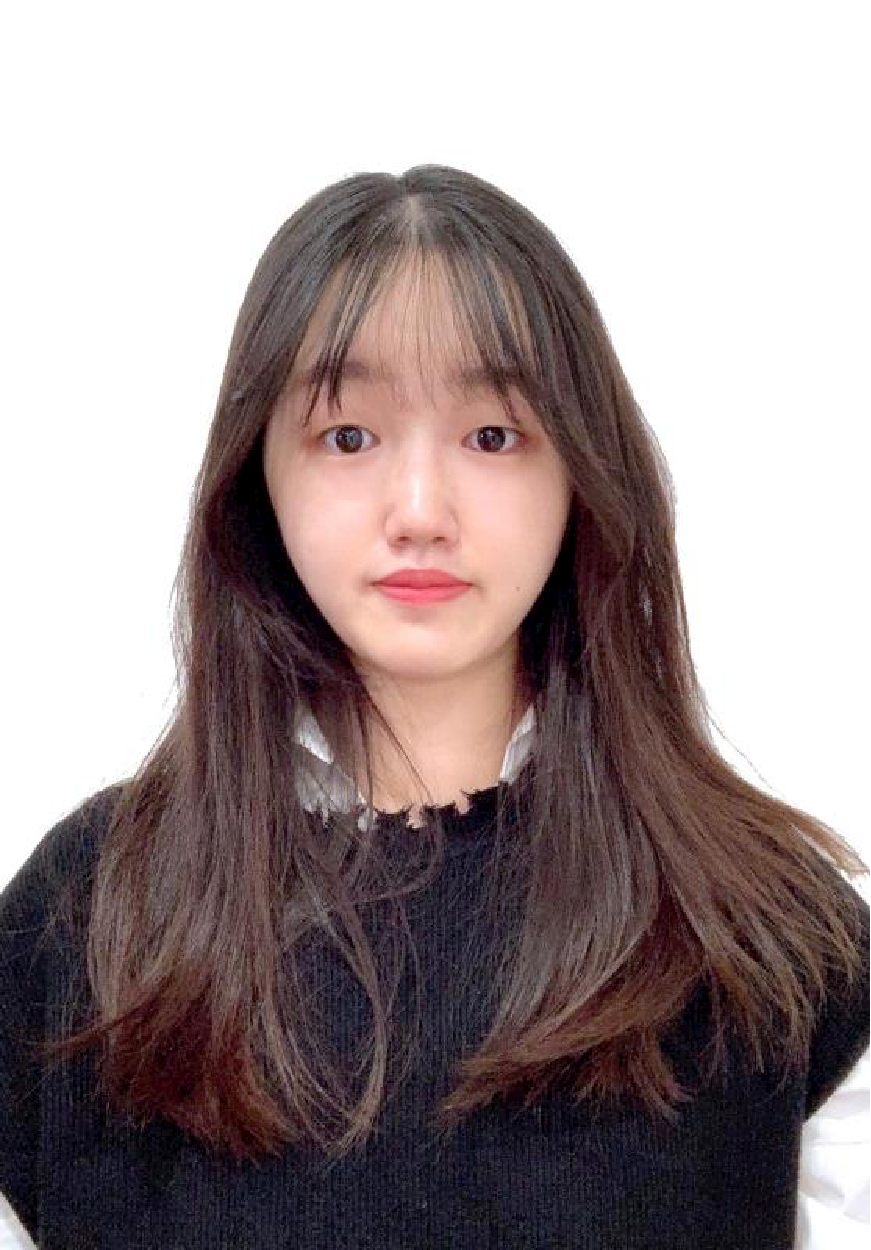}}]
{Xiaowen Cao}(S'18) received the B.Eng. degree from the Guangdong University of Technology, China, in 2017. She is currently pursuing the Ph.D. degree in the School of Information Engineering, Guangdong University of Technology, China. Her research interests include mobile edge computing and over-the-air comnputation.
\end{IEEEbiography}

\begin{IEEEbiography}[{\includegraphics[width=1in,height=1.25in,clip,keepaspectratio]{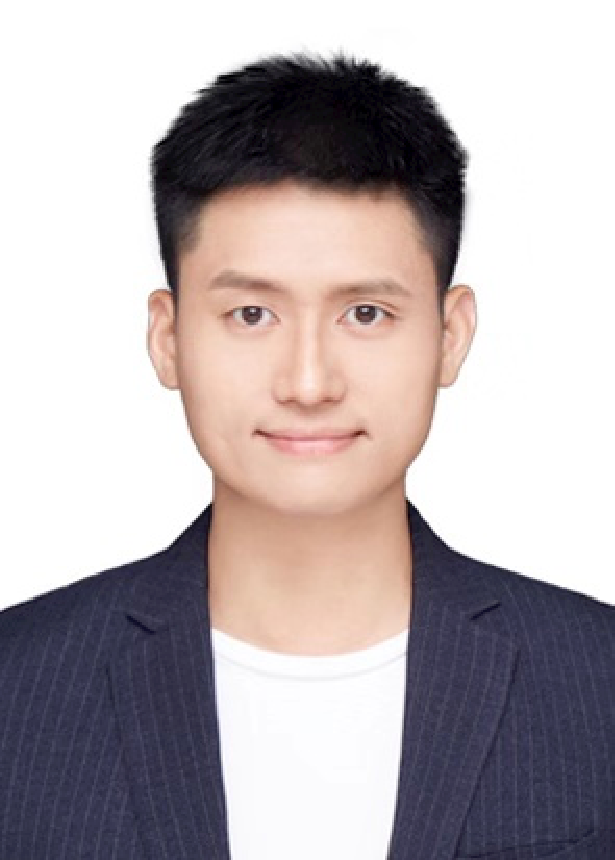}}]{Guangxu Zhu}
(S'14) received the B.Eng. and M.Eng. degrees from Zhejiang University, and the Ph.D. degree from The University of Hong Kong in 2019. He is now a research scientist with the Shenzhen Research Institute of Big Data. His research interests include edge intelligence, distributed machine learning, 5G technologies such as massive MIMO, mmWave communication, and wirelessly powered communications. He is a recipient of the Hong Kong Postgraduate Fellowship (HKPF) and a Best Paper Award from WCSP 2013.
\end{IEEEbiography}

\begin{IEEEbiography}[{\includegraphics[width=1in,height=1.25in,clip,keepaspectratio]{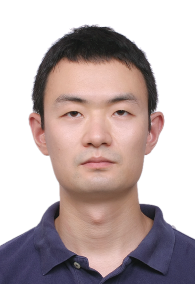}}]
{Jie Xu} (S'12-M'13) received the B.E. and Ph.D. degrees from the University of Science and Technology of China in 2007 and 2012, respectively. From 2012 to 2014, he was a Research Fellow with the Department of Electrical and Computer Engineering, National University of Singapore. From 2015 to 2016, he was a Post-Doctoral Research Fellow with the Engineering Systems and Design Pillar, Singapore University of Technology and Design. From 2016 to 2019, he was a Professor with the School of Information Engineering, Guangdong University of Technology, China. He is currently an Associate Professor with the School of Science and Engineering, The Chinese University of Hong Kong, Shenzhen, China. His research interests include energy efficiency and energy harvesting in wireless communications, wireless information and power transfer, UAV communications, and mobile edge computing and learning. He was a recipient of the 2017 IEEE Signal Processing Society Young Author Best Paper Award, the IEEE/CIC ICCC 2019 Best Paper Award, the 2019 IEEE Communications Society Asia-Pacific Outstanding Young Researcher Award, and the 2019 Wireless Communications Technical Committee Outstanding Young Researcher Award. He is the symposium co-chair for IEEE GLOBECOM 2019 Wireless Communications Symposium, the workshop co-Chair for IEEE ICC 2018 and 2019 Workshop on UAV Communications, and the tutorial co-chair for IEEE/CIC ICCC 2019. He served or is serving as an editor for IEEE Wireless Communications Letters and Journal of Communications and Information Networks, an associate editor for IEEE Access, and a guest editor for IEEE Wireless Communications, IEEE Journal on Selected Areas in Communications, and Science China Information Sciences.
\end{IEEEbiography}

\begin{IEEEbiography}[{\includegraphics[width=1in,height=1.25in,clip,keepaspectratio]{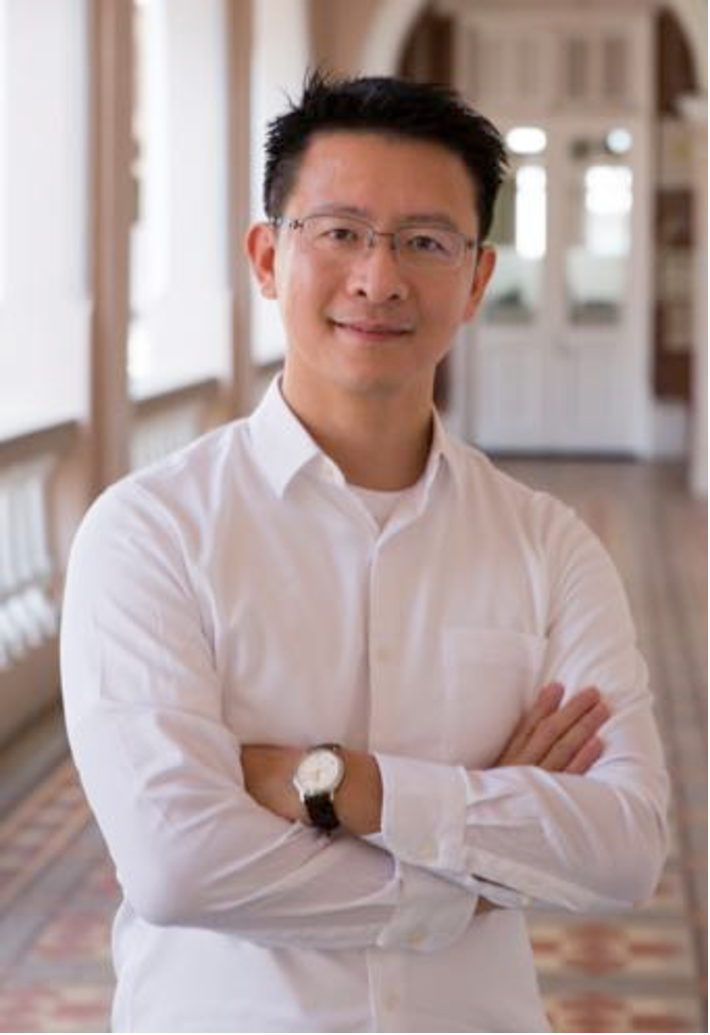}}]{Kaibin Huang}
(S'05-M'08-SM'13) received the B.Eng. and M.Eng. degrees from the National University of Singapore, and the Ph.D. degree from The University of Texas at Austin, all in electrical engineering. Presently, he is an associate professor in the Dept. of Electrical and Electronic Engineering at The University of Hong Kong. He received the IEEE Communication Society's 2019 Best Tutorial Paper Award, 2015 Asia Pacific Best Paper Award, and 2019 Asia Pacific Outstanding Paper Award as well as Best Paper Awards from IEEE GLOBECOM
2006 and IEEE/CIC ICCC 2018. Moreover, he received an Outstanding Teaching Award from Yonsei University in S. Korea in 2011. He has served as the lead chairs for the Wireless Comm. Symp. of IEEE Globecom 2017 and the Comm. Theory Symp. of IEEE GLOBECOM 2014 and the TPC Co-chairs for IEEE PIMRC 2017 and IEEE CTW 2013. He has edited special issues for IEEE JOURNAL ON SELECTED AREAS IN COMMUNICATIONS, IEEE JOURNAL ON SELECTED TOPICS IN SIGNAL PROCESSING, and IEEE Communications Magazine. He is/was an Associate Editor for several major journals in wireless including IEEE TRANSACTIONS ON WIRELESS COMMUNICATIONS, IEEE WIRELESS COMMUNICATIONS LETTERS, and IEEE TRANSACTIONS ON GREEN COMMUNICATIONS AND NETWORKING. He was an ISI Highly Cited Researcher in 2019.
\end{IEEEbiography}

\end{document}